\documentclass[aps,pra,twocolumn,showpacs,preprintnumbers,amsmath,amssymb,footinbib]{revtex4}
\usepackage{graphicx,epsfig}
\usepackage{bm}
\usepackage{dcolumn}
\usepackage{xcolor}% coloured text
\usepackage[breaklinks=true,colorlinks,citecolor=blue,linkcolor=blue,urlcolor=blue]{hyperref}

\def\i{\indent}
\def\noi{\noindent}
\def\bc{\begin{center}}
\def\ec{\end{center}}
\newcommand{\bea}{\begin{equation}}
\newcommand{\eea}{\end{equation}\noi}
\newcommand{\ber}{\begin{eqnarray}}
\newcommand{\eer}{\end{eqnarray}\noi}
\begin{document}
\title{Molecular dynamics study of nanoconfined TIP4P/2005 water: how confinement and temperature affect diffusion and viscosity}
%\author{Tanmoy Mondal}
%\author{Shyamal Biswas}\email{sbsp [at] uohyd.ac.in}
%\affiliation{School of Physics, University of Hyderabad, C.R. Rao Road, Gachibowli, Hyderabad-500046, India}

\author{A Zaragoza$^{1,2}$}
\author{MA Gonzalez $^{3}$}
\author{L Joly$^{4}$}
\author{I Lopez-Montero$^{3}$}
\author{MA Canales$^{5}$}
\author{AL Benavides$^{2}$}
\author{C Valeriani$^{1}$}
\affiliation{
$^{1}$Departamento de Estructura de la Materia, Facultad de Ciencias F\'{i}sicas, F\'{i}sica T\'{e}rmica y Electr\'{o}nica, Universidad Complutense de Madrid, 28040 Madrid, Spain. \\
$^{2}$ Depto. Ingenier\'{i}a F\'{i}sica, Divisi\'{o}n de Ciencias e Ingenier\'{i}as,Universidad de Guanajuato, 37150 Le\'{o}n, Mexico. \\
$^{3}$ Departamento de Qu\'{i}mica F\'{i}sica, Facultad de Ciencias Qu\'{i}micas,
Universidad Complutense de Madrid, 28040 Madrid, Spain. \\
$^{4}$Univ. Lyon, Universit\'{e} Claude Bernard Lyon 1, CNRS, Institut Lumi\`{e}re Mati\`{e}re,F-69622 Villeurbanne, France. \\
$^{5}$Instituto de Investigaci\'{o}n Hospital Doce de Octubre (i+12), Avenida de C\'{e}rdoba s/n, 28041 Madrid, Spain. \\
$^{6}$Departamento de Qu\'{i}mica Org\'{a}nica I, Facultad de Ciencias Qu\'{i}micas,
Universidad Complutense de Madrid, 28040 Madrid, Spain.}

\date{\today}

\begin{abstract}
In the last decades a large effort has been devoted to the study of water confined in hydrophobic
geometries at the nanoscale (tubes, slit pores), because of the multiple technological applications
of such systems, ranging from drugs delivery to water desalinization devices. To our knowledge,
neither numerical/theoretical nor experimental approaches have so far reached a consensual un-
derstanding of structural and transport properties of water under these conditions. In this work,
we present molecular dynamics simulations of TIP4P/2005 water under different hydrophobic
nanoconfinements (slit pores or nanotubes, with two degrees of hydrophobicity) within a wide
temperature range. On the one side, water is more structured near the hydrophobic walls, inde-
pendently on the confining geometries. On the other side, we show that the combined effect of
confinement and curvature leads to an enhanced diffusion coefficient of water in hydrophobic nan-
otubes. Finally, we propose a confined Stokes-Einstein relation to extract viscosity from diffusivity,
whose result strongly differs from the Green-Kubo expression that has been used in previous
work. We discuss the shortcomings of both approaches, which could explain this discrepancy.

\end{abstract}

%\pacs{03.65.-w, 03.65.Sq, 05.30.-d}

\maketitle
 
%\tableofcontents

\section{Introduction}

%\textcolor{red}{Laurent, could you suggest us how to mention the hydrodynamics in the introduction? }

Water is essential for life. Although it is formed by simple triatomic molecules, it has an unusual behavior when compared with similar mono-component substances.\cite{chaplin} It exhibits many anomalies and a lot of work has been devoted %in the scientific literature trying 
to understand and describe its anomalous behavior. Besides, under confinement, water exhibits a different phase diagram than bulk water, where the anomalies can disappear or occur at different thermodynamic conditions.\cite{man:cpl05} When confinement reaches the nanoscale, water undergoes novel physical properties, different from their bulk analog and still not completely understood.\cite{kog:letnat01,tak:pnas08, agr:nat17,cha:acs12,koh:cpl16}
Research on nano confined water has become of great interest due to its multiple nano technology applications, as for instance cells biochannels, \cite{Has:bio07,wam:bba17,taj:sci02} drugs delivering, \cite{Zha:dru10,Liz:Int17,Men:bio12,vas:car11} water desalination devices, \cite{wan:car17,cor:jpcb08,das:des14,get:app10,cor:ees11,bor:bul17,hum:nat01,fal:nanolett10} among others. \cite{alg:nat15,fal:nanolett10,bon:jpcm11,gal:conmat03,zan:prl03,zan:conmat04,zan:jcp03,gio:prl09,gio:pre06}

Due to the small length scales, the study of water in  nano confinement is experimentally very challenging, as shown by the few experimental works on the subject.~\cite{liu:langmuir14,man:cpl05,kol:prl04}
Besides, experimental results are controversial and open problems remain. \cite{has:nr16} This is the main reason why there has been a greater interest in using theoretical approaches or molecular simulations to understand, for instance, the effect of varying the geometrical parameters defining the nano confinement and/or the thermodynamic conditions. \cite{sud:acr17,Hin:jpc07,Li:prb07,ras:ann08,str:jcp05,bon:jpcm11,kro:jcp13,mos:jcp12,pug:jpcb17}
However, to our knowledge, neither the theoretical and simulations approaches, nor the experimental ones, have so far reached a consensus on how water transport properties are affected under these conditions.\cite{has:nr16,Ale:chemrev08}

Thus, molecular dynamics (MD) simulation is a very important tool to study confined water.~A good force field is then required to model the water-water
and the water-wall interactions.
For pure water simulation studies,
many force fields have been proposed to reproduce the complex phase
diagram of water and its anomalies;\cite{san:prl04} among them one can
list:
SPC,\cite{ber:int81} TIP3P,\cite{jor:jcp83} TIP4P,\cite{jor:jcp83} AMOEBA,\cite{ren:jcp03}
and TIP4P/2005,\cite{aba:jcp05} the latter being one of the most
robust and accurate. For confined water (between walls, inside pores
or nanotubes) using rigid atomistic models, most of simulation
works have been performed using TIP3P or SPC/E models and only  a few
have used the TIP4P/2005
model.~\cite{koh:pccp17,mar:ent17,sud:acr17,liu:jcp14,nak:mcp12,kum:pre05,ale:ces08,ale:ms08,bau:pre12,zhe:pcc12,gus:jpcb12,tsi:ms18,Gravelle2016,Fu2018}

A reliable MD simulation study for water transport in carbon nanotubes (CNTs) in order to find out conduction rates with different water models has been done by Liu and Patey \cite{liu:jcp14} with a pressure-driven water transport (non-equilibrium conditions) through (8,8) and (9,9) CNTs using three different water models (TIP3P, SPC/E and TIP4P/2005). In a second paper, the same authors have considered smaller tubes and rationalized the differences in terms of viscous entrance effects controlled by the viscosity of the water model at the entrance of the tube.~\cite{Liu2016a} 

Among recent simulation works on confined TIP4P/2005 water under
equilibrium conditions, one can mention the work of
K\"ohler et al. \cite{koh:pccp17} and Marti et al.\cite{mar:ent17}
Both groups considered this model either confined inside or outside CNTs
with hydrophobic or hydrophilic walls. Different diameters  were
considered and their effect on structural or transport properties were
analyzed for some selected thermodynamic conditions. They found
interesting results, however, since K\"ohler et al. studied CNTs
(10,10), (16,16), and (30,30), meanwhile, Marti et al. CNTs (5,5),
(9,9), and (12,12), and the latest authors used different water-carbon interaction
models, it is not possible to give a general conclusion of their
findings. For instance, K\"ohler et al. found a great influence of the
density, confinement size, and water-wall interaction on the
viscosity. 
Marti et al. \cite{mar:ent17} besides nanotubes, considered two graphene sheets separated by a distance from  0.6~nm up to 1.7~nm  immersed in TIP4P/2005 water at T~=~275K and p~=~400 bar. They found that the diffusion of water depends on the plates distance in a non-monotonic way. 
%They claim that this behavior is due to the water structuring, crystallization, re-melting and evaporation while the distance between plates is diminished.%
As can be noticed, even for this popular water model under confinement, the available information is limited and the same occurs with the experimental data.

So, in this work we try to cover this gap and have chosen the TIP4P/2005 water
 model confined between rigid hydrophobic carbon parallel walls, separated by a distance 
 1.6~nm  and 5.6~nm, and inside CNTs (20,20), (35,35) and (52,52). This study is carried out using MD simulations, underlying the effect (if any) played by the hydrophobicity of the surface and by the temperature of the system. Our study aims at being a benchmark for future works, explaining quantitatively the water structure under confinement, its dynamics and the number of hydrogen bonds per molecule formed inside these configurations. Additionally, in this work a few NMR experimental measurements to predict the diffusion coefficient of water in CNTs are presented and compared with available experimental data.

The manuscript is  organized as follows: 
In Section 1, the TIP4P/2005 water model and the water-carbon hydrophobic interactions are discussed. 
In Section 2, we present the simulation details as well as those of the nanostructures to be considered and implemented in simulations. 
In Section 3, simulation results for the structural properties (density profiles and hydrogen-bonds distributions) and transport properties (diffusion coefficient and viscosity) are given for different confinement geometries and temperatures. 
For viscosity, we derived a confined Stokes-Einstein formula, which we compare to the commonly used Green-Kubo formula. Finally, the main conclusions of this work are given in section 4.
This article is complemented with three  appendixes. 
%, one of them (Appendix-1) includes the calculus of the diffusion coefficient for a  prismatic nanotube to compare with the results obtained in this work for a cylindrical nanotube with roughly the same size. This study will allow to test whether the corrections for prismatic boxes are valid  for nanotubes. The other appendix 
Appendix A describes a drop-shape analysis to show how we selected the water-wall interaction. 
In Appendix B, a confined Stokes-Einstein relation is derived for a liquid confined between parallel stress-free walls, in the framework of continuum hydrodynamics.  
Appendix C contains our experimental attempt  of measuring the diffusion coefficient of water confined in CNTs using Nuclear Magnetic Resonance (NMR) technique.

\section{Systems and methods}
\label{sec:sim_details}
%\section{Nanostructures}
%We need to select and prepare the carbon nanostructures before carrying out the simulations.

Throughout this work, we have carried out NVT simulations using the Molecular Dynamics package GROMACS 2016.4,\cite{GROMACS}
for water confined under two geometries: between two parallel graphene walls (W) and inside carbon nanotubes (CNT). 
We made this study for several temperatures in the range 243~K to 298~K. 
We set the  timestep to 1~fs and simulate every temperature for at least 40~ns, and for lowest temperatures (243-253 K) up to 60~ns.
In order to keep the temperature constant, we use a Canonical sampling through velocity rescaling thermostat~\cite{bus:jcp07} with a relaxation time set to 1~ps. 
We checked that the same results were obtained by means of Berendsen or Nos\'e-Hoover thermostats.  %(data not shown). 
We are aware of the fact that thermostating confined fluids is a very delicate issue \cite{ber:jcp10,bab:jcp11,kan:jcp12,tao:molsim18} given that there is no perfect way to thermostat a confined liquid. However, what we have used in our work is not an uncommon approach. We have checked different damping times (1ps being a long relaxation time), and thermostats to make sure that the chosen thermostat was not affecting our results.

%\pb{As a referee, I would be worried by this thermostat: velocity rescaling algorithms are known to affect the dynamics more than other ones; in general thermostating a confined liquid can be problematic: have you made some tests on the impact of the thermostat? What is the exact name of the thermostat? I would be slightly more comfident with, e.g., a canonical sampling velocity rescaling thermostat. }
We used a block average method to estimate the error bars \cite{Hes:jcp02}.

\subsection{Tuning the hydrophobicity of the interaction potential}

Water molecules %(OO interactions) 
have been simulated using the TIP4P/2005 force field~\cite{aba:jcp05}. 
%a rigid non-polarizable interaction potential well known to 
% perform very well when studying  water and its anomalies \cite{aba:pccp11}.
 We truncated the Lennard-Jones (LJ) potential at 9.5~\AA, adding standard tail corrections to the LJ energy, and 
considered Ewald sums (with the PME technique) \cite{ess:jcp95} for the calculation of the long-range electrostatic forces, applying a real space cut-off at 9.5~\AA. 
%{\bf CV improve caption figure}
 %\footnote{
 %his potential considers interactions among oxygens (OO) through a Lennard-Jones potential and among charges through Coulomb interactions. 
%The parameters for this model are given in Table \ref{tab_pots}.} 
% \pb{To characterize carbon-carbon (CC) interactions, we have  used the OPLS-AA force field~\cite{jor:jacs88}. LAURENT: I would remove this sentence: since carbon atoms are fixed, C-C interactions are irrelevant. } 
The carbon-oxygen (CO) diameter interaction is described by a geometric average, %~\cite{Lor:adp81} 
% have been considered as a Lennard-Jones potential having a diameter described by $\sigma_{CO}$ according to the Lorentz-Berthelot rule~\cite{Lor:adp81}, 
%\begin{equation}
$\sigma _{CO} = \sqrt{\sigma _{CC}\sigma _{OO}}$, 
where $\sigma _{CC}$ is taken from the OPLS-AA force field~\cite{jor:jacs88}.    
%\end{equation}
The value of $\varepsilon_{CO}$ established the amount of hydrophobicity of the carbon wall.
In this work we have used a value of $\varepsilon_{CO}$ widely used for water in graphite confinement \cite{mos:jcp14,hum:nat01,wag:jcp02,koh:pccp17} correponding to a contact angle of around $96.5^{\circ}$. For comparison, we have also used a superhydrophobic water-carbon interaction value previouly used by Algara-Siller et al. in Ref. \cite{alg:nat15} corresponding to a higher contact angle, $132^{\circ}$ and $\varepsilon_{co}^{sh}$~=~0.0476~kJ/mol. % (see Fig.~\ref{drop1}-b).
%The LJ parameters for this cross interaction are collected in Table~\ref{tab_pots} from Appendix 1.
%All  parameters are reported in Appendix-1 (Table \ref{tab_pots}).  
%
%\noindent \large{\textbf{Appendix 1: Contact angle estimate}}
%
As detailed in Appendix A, the contact angles were estimated by simulating a sessile (nano)droplet of water on planar walls. 
In Table~\ref{tab_pots}, we report  the  parameters of the force fields used for water and carbon.
%Throughout the work, we have used two oxygen-carbon interactions, one hydrophobic, C-O$_{w}$ (H), and other superhydrophobic, C-O$_{w}$ (SH).

\begin{table}
\small
	\caption{TIP4P/2005 water model force field parameters. A  Lennard-Jones potential characterizes the oxygen-oxygen, carbon-carbon (OPLS-AA) and carbon-oxygen interactions 
	($\varepsilon_{co}^{h}$ and $\varepsilon_{co}^{sh}$). The charge (q$_{H}$ and q$_{M}$ are the hydrogen and dummy atom charges respectively) and angle refer to  TIP4P/2005 water model.}
	\begin{tabular*}{0.48\textwidth}{@{\extracolsep{\fill}}lll }
% \begin{tabular}{|l c c|}
  \hline
  \hline
	Interaction &  $\sigma$(nm) & $\varepsilon$ (kJ/mol)   \\
  \hline
	  \multicolumn{3}{c} { \textit{Lennard-Jones parameters}} \\
  \hline
	  O-O & 0.3159 &  0.7749 \\
	  C-C & 0.3550 &  0.2929 \\
	  C-O(h) & 0.3349  &  0.2703 \\
	  C-O(sh) & 0.3349  &  0.0472 \\
  \hline
	  \multicolumn{3}{c} {\textit{Charges (e)} } \\
  \hline
	  q$_{H}$ = -q$_{M}$/2 & ~ & 0.5564 \\
  \hline
  \hline
	  \multicolumn{3}{c} {\textit{H$_{2}$O angle (degrees) } } \\
  \hline
	  $\angle$ H$_{2}$O & ~ & 104.52$^{\circ}$ \\
  \hline
  \hline
\end{tabular*}
\label{tab_pots}
%\caption{Force field parameters of the TIP4P2005 water model under different geometrical confinements. Lennard-Jones parameters characterize the oxygen-oxygen interaction, carbon-carbon interaction,$\varepsilon_{co}^{h}$  and $\varepsilon_{co}^{sh}$ carbon-oxygen interaction.The charge and the geometry parameters are those of the TIP42005 model.}
\end{table}

\subsection{System preparation}

\begin{figure}
\begin{center}
\includegraphics[clip,angle=-90,width=0.5\textwidth]{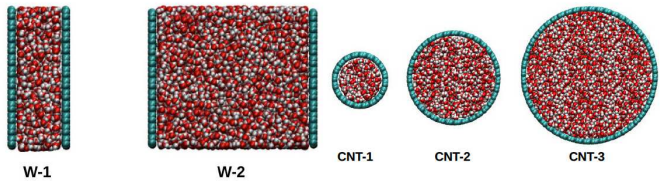}
	\caption{Snapshots of the nano-structures used in this work, made of  carbon atoms  (cyan spheres), filled with   water molecules (red  for oxygens and white  for hydrogens). 
	Confined systems between graphene walls are labeled as "W", and inside nanotubes as "CNT". Different numbers refer to systems with an increasing wall-distance (W) or diameter (CNT).}
\label{fig_schem-CNT}
\end{center}
\end{figure}

To start with, we prepare the general geometries filled with water molecules as shown in  
	Fig.~\ref{fig_schem-CNT}. Carbon atoms are placed on a lattice of graphene with fixed positions. Concerning nanotubes, atoms are placed according to the chirality specified in each case, also with fixed positions (frozen). The Carbon-Carbon interaction parameters ($\varepsilon$ and $\sigma$) are presented in table \ref{tab_pots}. These values correspond to OPLS-AA force-field as it was reported in reference \cite{jor:jacs88}.

%shows the nanostructures and its labels performed to study confined water under different confinements: between two parallel planes and within cylindrical confinement.

The W geometry consists of 
%The wall-confinement has been carried out performing
 two 5.2~nm$\times$5.2~nm parallel graphene walls, each made of 1008 carbon atoms  (1 atom thick).
 %at a distance of 5.6~nm. 
 In order to be able to apply periodic boundary conditions in all directions, the system, {\it i.e.} water and carbon walls, is located at  the center of the simulation box 
and an empty region is left 
%keeping an empty region 
between the parallel walls and the system's periodic replicas
%, since periodic boundary conditions have been applied in all directions. 
(as suggested  in Ref.~\citenum{str:jcp05}).
We  prepare two setups: W-1, containing 1361 water molecules confined between hydrophobic walls located at 
a distance of  1.6~nm; and W-2 containing 5417 water molecules confined between hydrophobic walls located at 
a distance of 5.6~nm.

%For carbon nanotubes, the geometry of the graphene lattice and chiral vector of the tube determines its structural parameters such as diameter, unit cell and  number of Carbon atoms \cite{nanot04}. 
%The diameter of the tube is given by the length of the chiral vector:
%\begin{equation}
%d = \frac{\left |c \right |}{\pi} = \frac{a_{0}}{\pi} \sqrt{n^{2}_{1} +n_{1}n_{2}+n^{2}_{2} }
%\end{equation}
%where $c$ (defined in terms of the pair of integers (n$_{1}$ and n$_{2}$)) uniquely defines a particular tube.
%\subsubsection{Carbon nanotubes.}
The CNT geometry consists of cylindrical hydrophobic nanotubes of the following chirality:  
%confinement we perform a nanotube.  
%The chirality of nanotubes from left to right is: 
(20,20) (CNT-1), (35,35) (CNT-2) and (52,52) (CNT-3). 
% Every system as well as the number of molecules inside are described in Table~\ref{tab_CNT-dim}
We use the Visual Molecular Dynamics software~\cite{HUMP96} to prepare every nanotube. % setting the length as well as the chirality of the tube. 
A single wall CNT is centered in the middle of the simulation box.
%avoiding replicas interacting along the X and Y direction.
Periodic boundary conditions are applied in all directions, but the interactions between the system and its replicas along X and Y are avoided by using very large simulation box sizes as it was shown by Zheng et al.~\cite{zhe:pcc12}
Numerical details of these setups are reported in  Table~\ref{tab_CNT-dim}.

%%%%%%%%%%%%%%%%%%%%%%%%%%%%%%%%%%%%%%%%%%%%%%
\begin{table}
	\caption{Top: Numerical details for the W-type confinement: number of carbon atoms per wall (N$_{C}$), distance between parallel walls (d), wall's edge (corresponding to the simulation box limit, L$_\text{Box}$) and number of water molecules (N$_{W}$). 
Bottom: Numerical details for the CNT-type confinement: chirality,  number of carbon atoms per nanotube, nanotube diameter, nanotube length (corresponding to the simulation box edge) and number of water molecules.} 
%Up Table: dimensions, chirality of the CNT and number of water molecules (N$_{w}$) inside. $L_{Box}$ represents the box length used to calculate the density. 
%Down table: dimensions of the planar systems and number of water molecules (N$_{w}$) inside.
%*CNT-0 has been thoroughly studied in Appendix-4.}
	\label{tab_CNT-dim}
	 \begin{tabular*}{0.48\textwidth}{@{\extracolsep{\fill}}lcccc}
	\hline
\hline
		 System & $d$(nm) & $L_{Box}$(nm) &N$_{C}$(per wall)  & N$_{w}$ \\
\hline
		 W-1  & 2.0 & 5.1 & 1008  & 1361 \\
		 W-2  & 6.1 & 5.2 & 1008 & 5417 \\
\hline
\hline
	\end{tabular*}	
		\vspace{0.1cm}
	\\
  \begin{tabular*}{0.48\textwidth}{@{\extracolsep{\fill}}lccccc}
	  %\hline

%%%%%%%%%%%%%\begin{table}[ht!]
%%%%%%%%\begin{center}
%%%%%%%%%%%%%%\hrulefill 

%%%%	\begin{tabular}{|c|c|c|c|c|c|}
\hline 
\hline 
	System & $(n,m)$ & $d$(nm) & $L_{Box}$(nm) &   N$_{C}$  & N$_{w}$ \\
\hline 
	CNT-1	&(20,20) & 2.6    &  5.16   & 1680  &  640  \\
	CNT-2	&(35,35) &  4.7    &  5.15 & 2940   &  2242 \\
	CNT-3	&(52,52) &  7.0    &  6.20  & 5201   &  6407 \\
\hline
\hline 
\end{tabular*}
%%%%%\quad
%%%%%%\hrulefill	
%	\begin{tabular}{|c|c|c|c|c|}
	
%\end{center}
%%%%%	\caption{Left Table: Dimensions and chirality of the CNT and number of water molecules (N$_{w}$) inside. $L_{Box}$ represents the box length used to calculate the density. Right table: Dimensions of the planar systems and number of water molecules (N$_{w}$) inside.}

	%The chirality is represented by (n,m). In this study we have considered armchair chirality (n=m). The nanotube diameter is $d$ and $L$ corresponds to the length of the carbon nanotube, both longitudes expressed in nm. The diameter of a nanotube is obtained by the expression: $d=\frac{a}{\pi} \sqrt{n^{2}+nm+m^{2}}$, where $a=0.246$. {\bf CV: WHERE DOES THIS EXPRESSION COME FROM? WE NEED TO ADD A REFERENCE. AND WHY a HAS THIS VALUE?}
%%%\label{tab_CNT-dim}
\end{table}

%The dimensions for the systems used in this work and the number of water molecules confined in those systems are collected in Table~\ref{tab_CNT-dim}.

%In nanotubes, the graphene sheet is rolled up in such a way that a graphene lattice vector c = n$_{1}$a$_{1}$+n$_{2}$a$_{2}$ becomes the circunference of the tube. This circunferential vector "c" which is usually denoted by the pair of integers (n$_{1}$ and n$_{2}$), is called the chiral vector and uniquely defines a particular tube.
%The direction of the chiral vector is measured by the chiral angle $\Theta$, which is defined as the angle between a$_{1}$ and "c":
%\begin{equation}
%\cos \Theta = \frac{a_{1} \cdot c}{\left |a_{1}  \right |\cdot\left|  c \right |} = \frac{n_{1} + n_{2}/2 }{\sqrt{n^{2}_{1} + n_{1}n_{2} + n_{2}} }
%\end{equation}	

%%%%%%%%%%%%%%%%%%%%%%%%%%%

To establish the density of confined water, we compute it 
as the number of water molecules divided by the volume that they are contained in, N$_{w}$/V 
($V$ being either the volume between the two walls or the volume inside the nanotube). 
	A recipe to estimate the volumes for the two geometries will be given in the Results section. For ultra-confined systems this could make a difference as compared to fixing the pressure. Whereas here the systems are above the critical size where sub-continuum behaviour is expected.~\cite{tho:prl09}

\subsection{Measured quantities}

In order to give a molecular explanation of water behavior under nano-confinement, we firstly computed the density profile of water molecules 
and calculated 
 the average number of hydrogen bonds (HBs) per molecule.
 %  considering the hydrogen bond distance explained in ref. ~\cite{kum:jcp07} and
 %> angle necessary to bond two water molecules d $\approx$~0.28~nm and $\widehat{OH} = 109.4^{\circ}$. 
  The HB criterion is based on the number of donor hydrogen-bonds per molecule as proposed by Kumar et al.~\cite{kum:jcp07}, where 
   the HB distance and angle necessary to bond two water molecules are $d \approx 0.28$\,nm and $\widehat{OH} = 109.4^{\circ}$.

Next, we computed the viscosity of confined water which can be a delicate issue in confined systems.~\cite{wan:amss11,mor:jcp17,han:lan15,dai:pro18,hoa:pre12}

In a bulk homogeneous liquid, viscosity can be computed by means of the Green-Kubo expression:
\begin{equation}
\eta_\text{GK} = \frac{V}{k_{B}T} \int^{\infty}_{0}dt \left \langle P_{\alpha 
\beta}(t)P_{\alpha \beta}(0) \right \rangle ,
\label{eq:visco}
\end{equation}
% as the integral of the autocorrelation function of the components of the pressure tensor, calculated 
where P$_{\alpha\beta}$ are the traceless components of the pressure tensor. 
%
%In a confined liquid however, the system becomes heterogeneous and the Green-Kubo formula provides only an effective viscosity -- resulting from an average of viscous stress in the liquid and friction stress on the walls -- whose physical meaning is unclear. 
In a confined liquid however, the system becomes heterogeneous, and more important, the thermal fluctuations of the shear stress at equilibrium will not only be affected by the liquid viscosity, but by the slip boundary condition at the wall; consequently, the Green-Kubo formula provides only an effective viscosity whose physical meaning is unclear. 
Nevertheless, it can be interesting to compute $\eta_\text{GK}$ and to compare it with other possible measurements of confined viscosity, as discussed later.  
Note also that in confined systems, the volume $V$ of the liquid is not well defined, with a related uncertainty on the computed $\eta_\text{GK}$, and that  the stress tensor components $P_{\alpha \beta}$ depend on the system's geometry. 
Therefore, we must distinguish between axial viscosity (defined 
by axial pressure components P$_{xy}$) and radial viscosity (defined by radial 
components P$_{xz}$ and P$_{yz}$), as proposed by K\"olher et al.  
In this work, we have calculated for the two different types of confinement
 the axial viscosity, integrating the autocorrelation function of the axial 
components.
% for CNT and for walls has been computed.  

Another approach to estimate the viscosity relies on computing the self-diffusion coefficient and using a confined Stokes-Einstein relation.
% 
%Finally, we compute the long time diffusion coefficient 
% It is necessary to know the viscosity  in order to have a correct estimate of the water diffusion  coefficient under confinement. 
% ($D$) 
 %for every system studied we follow 
%the Eq.~\ref{diff} (general expression for 3D) particularising for diffusion in 
%1 or 2 dimensions,
 %where $D$ is related to the time and
The diffusion coefficient $D_{||}$ under confinement can be measured using the particle's mean square displacement at long time: %\textcolor{blue}{AL corrections not implemented} 
\begin{equation}
D_{||}= \frac{\left \langle | r(t_{0}+t) -r(t_{0}) |^{2}  \right  \rangle }{ 2\, {\it dim}\times  t}
\label{diff}
\end{equation}
where ${\it dim}$ depends on the system's geometry: being ${\it dim=1}$ for CNT and ${\it dim=2}$ for parallel walls.

Diffusion measurements in bulk liquids via MD are strongly affected by finite size effects due to hydrodynamic interactions with periodic images of the simulation box \cite{kre:jcp93,yeh:jpc04,pau:jctp17,mar:jpcb16}. 
%To compute the "real" diffusion of a bulk system with periodic boundary conditions  in simulation studies one should  include corrections due to the finite size of the simulation box.
Analytical corrections exist both for isotropic or anisotropic simulation boxes. 
However, the case of confined systems is quite different and has been less explored.
In planar confinement, assuming a no-slip boundary condition on the walls, Simonnin et al.\cite{Simonnin2017} have computed analytically the effect of liquid height $d$ and box lateral size $L$ on the diffusion coefficient. 
Here we would like to emphasize that, while the effect of the finite lateral size $L$ is purely a limit of the simulation, the confinement height $d$ has a real physical effect. Indeed, diffusion is affected in the vicinity of walls. 
Analytical descriptions can be found for planar confinement following Faxen's pioneering work \cite{HappelBrennerBook}; in particular, it has been shown that liquid-solid slip impacts how diffusion is affected close to walls.~\cite{Saugey2005,Lauga2005,Joly2006a}
Regarding the effect of the lateral size $L$, the exact formula derived by Simonnin et al. \cite{Simonnin2017} can be approximated by assuming that the particle diffusion coefficient is the sum of their intrinsic diffusion coefficient and the diffusion coefficient of the liquid center of mass (com), for which analytical predictions exist in the presence of liquid-solid slip.~\cite{Detcheverry2012,Detcheverry2013} In practice, this also means that removing the liquid com motion when computing the particle mean square displacement -- as we did in this work -- provides a good estimate of the intrinsic diffusion coefficient for infinite lateral size $L$. Then only the physical effect of the confinement remains. Simonnin et al. derived an expression for the average parallel diffusion coefficient $\langle D_\| \rangle$ in the no-slip case, using Einstein relation between mobility and diffusivity \cite{Einstein1905}, and a continuum hydrodynamics description of Stokes drag in confinement. In graphite confinement, slip is very large, with slip lengths typically much larger than the confinement.~\cite{Holt2006,Secchi2016,Kannam2013,Falk2010} It is therefore reasonable to assume a perfect slip boundary condition (stress-free, infinite slip length). Under those conditions, and for moderate planar confinement, one can adapt the calculation by Simonnin et al. (see Appendix B) to show that: 
\begin{equation}\label{eq:confinedSE}
	\langle D_\| \rangle \approx \frac{k_\text{B}T}{3 \pi \eta \sigma_\text{h}} \left[1 + \frac{3}{8} \frac{\sigma_\text{h}}{d} \ln  \left( \frac{2d}{\sigma_\text{h}} \right) \right] , 
\end{equation}
with $\eta$ the viscosity (assumed homogeneous and isotropic) and $\sigma_\text{h}$ the effective hydrodynamic diameter of the particles \cite{pab:jcp18}, which can be computed from bulk measurements of diffusion and viscosity (see Appendix B). In the following, we shall refer to Eq.~\eqref{eq:confinedSE} as the confined Stokes-Einstein relation. 
Regarding CNT systems, we are not aware of an expression equivalent to Eq.~\eqref{eq:confinedSE} in cylindrical geometry. As a very rough estimate, we suggest to use the same equation, replacing the slab height by the tube diameter. The confined Stokes-Einstein relation provides an alternative estimate of the viscosity: 
\begin{equation}\label{eq:eta_SE}
\eta_\text{SE} \approx \frac{k_\text{B}T}{3 \pi \sigma_\text{h} \langle D_\| \rangle} \left[1 + \frac{3}{8} \frac{\sigma_\text{h}}{d} \ln  \left( \frac{2d}{\sigma_\text{h}} \right) \right] , 
\end{equation}
which we will compare to the Green-Kubo estimate in the results section. 
Equation~\eqref{eq:eta_SE} provides an effective viscosity, averaged in space and over the different traceless components of the pressure tensor; however, unlike with the Green-Kubo formula, here the effect of the slip boundary condition on the walls is taken into account and separated from the intrinsic viscosity of the confined liquid. 
%Note that Eq.~\eqref{eq:eta_SE} represents the first order of a Taylor expansion in the limit $\sigma_\text{h} \ll d$; for very strong confinement, the full analytical solution could be used instead. 
Note that the Stokes-Einstein relation is known to break down in supercooled bulk water \cite{Guillaud2017,pab:jcp18}, so that the confined Stokes-Einstein estimate of the viscosity should be taken with caution at very low temperature.

\section{Results}
\label{sec:results}

%Confinement may modify the orientation and configuration of water molecules inside the structure as well as near the walls~\cite{ala:chemphys11}. 

In this section we present our results on water structural (density and HBs) and transport (viscosity and diffusion) properties. 
%computed and analyse the most relevant properties of water, such as density and hydrogen
 %bonds (HB) profiles, viscosity and diffusion, by measuring the structural properties, which have molecular information about the confined water behaviour.  
  We study the effect 
   of hydrophobicity, of nanostructure's curvature (comparing  parallel and cylindrical confinement) 
    %comparing between wall-confinements and among cylindrical structures
 and of temperature on each of the structural and transport properties. 
%  and comparing between different type of structures -walls and cylindrical confinements-, 
% and finally, we study the trend of the properties with temperature.

\subsection{Density profiles}

We computed the density profiles of confined water for every system at several temperatures for both hydrophobic and superhydrophobic 
interactions.
As already observed, the density profile of confined water differs with respect to bulk water, because of  
 curvature effects and water-carbon interaction.~\cite{fal:nanolett10,cui:ija15} 

\begin{figure}
\begin{center}

	\includegraphics[width=0.7\linewidth]{comparoepsilon-35x35_v2.eps}a)\\[2mm]
	\includegraphics[height=1.75cm]{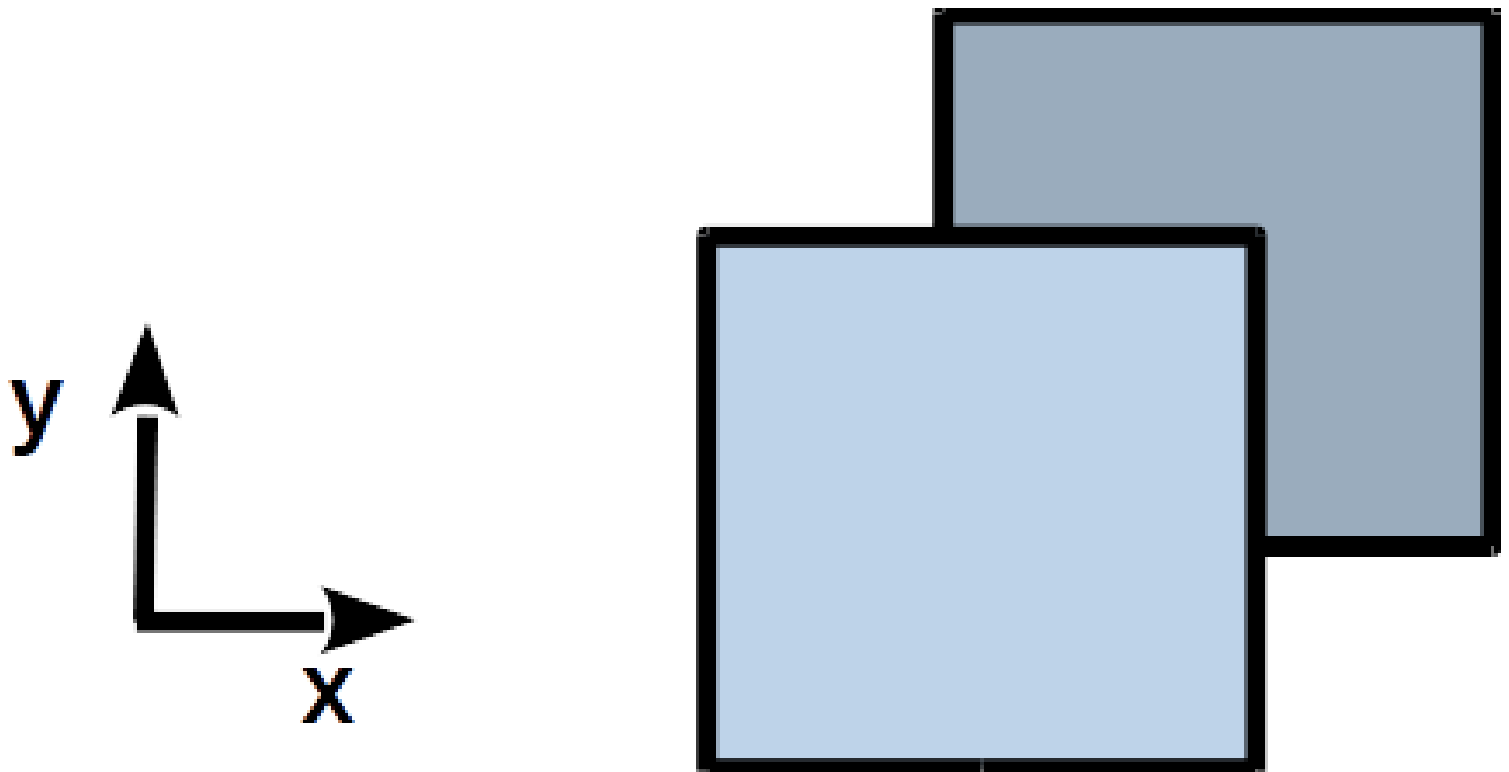}b) 
	\hspace{4mm} \includegraphics[height=1.75cm]{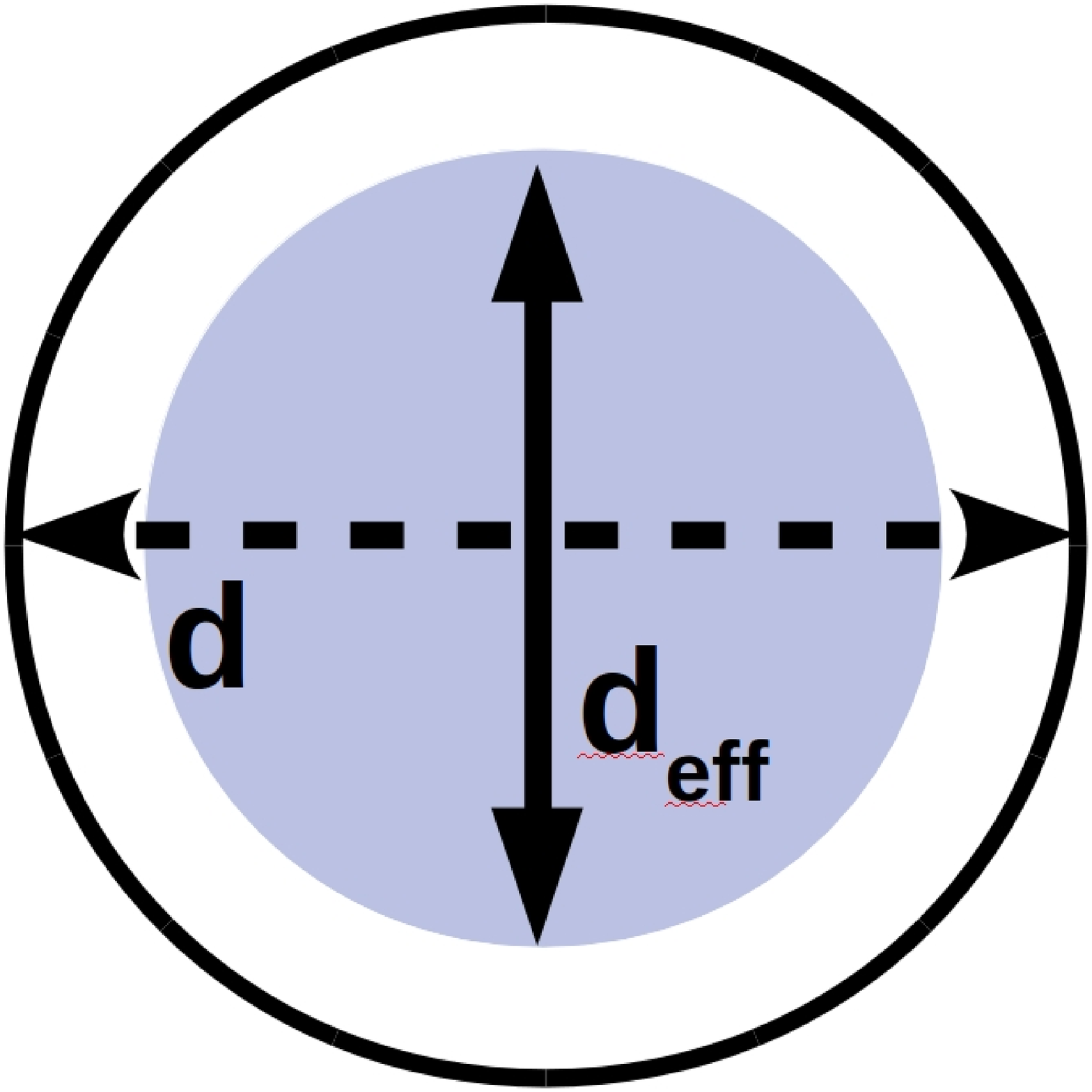}c)

\caption{a) Zoom close to the hydrophobic wall ($\varepsilon_{co}^{h}$) (purple vertical lines) of the 
water density profiles (red curve) computed from the center of a CNT-2 nanotube at 
$T = 298$\,K. The meaning of the presented variables is reported in the text. 
b) and c) are schematic representations of the orientations of both parallel walls (b) and  nanotubes (c). 
In the latter, $d$ indicates the nominal diameter, whereas $d_{eff}$ indicates the effective one.}
%Water diffuses in \textit{x} and 
%\textit{y} directions for walls (b) while nanotubes (c) are oriented along the 
%\textit{z} axis. }
\label{walls}
\end{center}
\end{figure}

As shown in Fig.~\ref{walls}, when preparing a CNT-2 hydrophobic system containing liquid water 
with a nominal density of 1\,g\,cm$^{-3}$, the density profile reveals water layering, especially close to the 
hydrophobic walls. The hydrophobicity of the confining walls expels water molecules of the closest layer, thus 
 %This effect generates an empty region between the first water layer and the wall, 
reducing the overall confining volume available to the water, which leads to an "effective density" that is higher 
than the nominal one. %, being smaller the available volume. 
To compute the effective volume for the walls and for the nanotubes, the effective distance $d_{eff}$
% molecules thus increasing 
%the "effective density" of confined water, which is the first aspect that we have to consider to give the proper density of confined water.
%Thus, we estimate the effective density of confined water between parallel walls 
%or in nanotubes.
is defined as $d_{eff}=d-\sigma_{CO}-2d_{0}$, 
where $d$ is the nominal wall-to-wall distance.  
%$d_{0}$ is a parameter that depends 
%on  the surface, being hydrophobic or superhydrophobic, 
%%a higher repulsive  interaction implies that the effective volume will be smaller than in the hydrophobic case.
%multiplied by 2 for the symmetry of the confinement (being the emptier region 
%symmetrical near the walls). 
To estimate $d_{0}$, we first compute the density profile from the center of 
the structure (whether parallel walls or nanotube) to the wall. 
Then, we calculate the distance between the $Z$ value (for planar confinements) 
or $r$ value (for cylindrical structures)  at the highest density ($d(\rho=max)$) and $Z$ or $r$ 
position at zero density ($d(\rho=0)$, as shown in  Fig.~\ref{walls}).  
Finally diving by two (for symmetry reasons), we compute 
%This length is divided by 2 and from this position, 
$d_{0}$ as
\begin{equation}
d_{0} =  \frac{1}{2} \sigma_{CO} - \frac{1}{2} \left( |d(\rho = max) - d(\rho = 0)| \right)
\end{equation}

The effective distance ($d_{eff}$), volume 
($V_{eff}$) and density ($\rho_{eff}$), for all the systems considered in this work are reported in Table~\ref{deff-w}.

\begin{table}
	\caption{Nominal distance between walls/the nanotubes diameters $d$ , 
	and their effective values $d_{eff}$ , effective volume $V_{eff}$ and effective density $\rho_{eff}$
(taking the number of water molecules for each system from Table~\ref{tab_CNT-dim}).}
\label{deff-w}
 \begin{tabular*}{0.48\textwidth}{@{\extracolsep{\fill}}lcccc}
		%\begin{tabular}{|c | c |c | c| c|}
\hline
System& 
$d$(nm)& 
$d_{eff}$(nm) & $V_{eff}$($nm^{3}$) & $\rho_{eff}$(g$\cdot cm^{-3}$) \\
\hline
%\multicolumn{4}{c} {\textbf{$\varepsilon_{co}^{sh}$}} \\
\hline
W-1 \textbf{$\varepsilon_{co}^{sh}$} & 2.00  & 1.50 & 38.8 & 1.1 \\
W-2 \textbf{$\varepsilon_{co}^{sh}$}  & 6.10 & 5.60 & 151.4 & 1.1 \\
%\hline
%\multicolumn{4}{c} {| \textbf{$\varepsilon_{co}^{h}$} |} \\
\hline
W-1 \textbf{$\varepsilon_{co}^{h}$} & 2.00 &1.50 & 38.9 & 1.0 \\
W-2 \textbf{$\varepsilon_{co}^{h}$}  & 6.10 &5.60 & 151.4 & 1.1 \\
\hline
\hline
		\vspace{0.01cm} \\
\hline
%\multicolumn{4}{c} { \textbf{ $\varepsilon_{co}^{sh}$  }} \\
\hline
CNT-1 \textbf{$\varepsilon_{co}^{sh}$}  & 2.60 & 1.97 & 15.73 & 1.2 \\ 
CNT-2 \textbf{$\varepsilon_{co}^{sh}$}  & 4.74 & 4.05 & 66.24 & 1.0 \\
CNT-3 \textbf{$\varepsilon_{co}^{sh}$}  & 7.04 & 6.28 & 192.04 & 1.0 \\
%\hline
%\multicolumn{4}{c} { \textbf{ $\varepsilon_{co}^{h}$ }} \\
\hline
CNT-1 \textbf{$\varepsilon_{co}^{h}$}  & 2.60 & 2.02 & 16.02 & 1.2 \\
CNT-2 \textbf{$\varepsilon_{co}^{h}$}  & 4.74 & 4.20 & 71.23 & 0.9 \\
CNT-3 \textbf{$\varepsilon_{co}^{h}$}  & 7.04 & 6.36 & 196.97 & 1.0 \\
\hline
\hline
\end{tabular*}
\end{table}

\begin{figure}
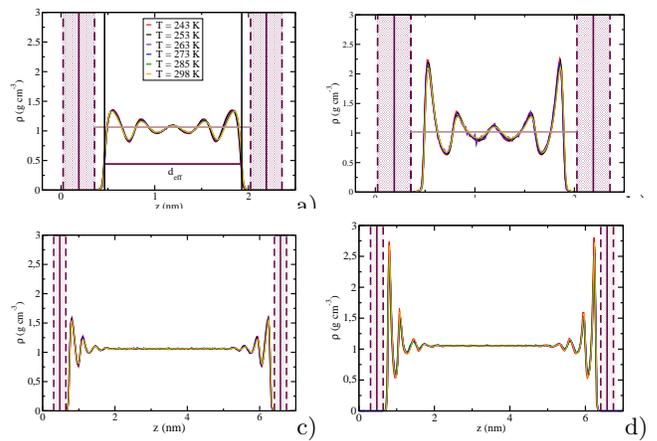

\begin{center}
\includegraphics[width=0.44\linewidth]{density-ALL_0047-smallwalls.eps}a)
\includegraphics[width=0.46\linewidth]{density-ALL-smallwalls-027.eps}b)\\
\includegraphics[width=0.44\linewidth]{densprof-newwalls-0047-5417.eps}c)
\includegraphics[width=0.46\linewidth]{new-walls5417-densprof-027.eps}d)
\end{center}
\caption{Density profiles  for water confined in W-1 (top panels) and W-2 (bottom panels) geometries at different temperatures (legend shown in figure a)). 
	 Results for  the superhydrophobic system $\varepsilon_{co}^{sh}$ are on the left-hand side (a) and (c)
	whereas for the hydrophobic system $\varepsilon_{co}^{h}$ are on the right-hand side (b) and (d) .}
\label{dens-walls}
\end{figure}

Figure~\ref{dens-walls} represents the   
density profile obtained for confined water under planar confinement (W-1 and W-2) in a wide temperature range,  considering the hydrophobic and superhydrophobic walls.
 The top panels of Fig.~\ref{dens-walls} are the results for the W-1  superhydrophobic (Fig.\ref{dens-walls}-a) and hydrophobic (Fig.\ref{dens-walls}-b) slab geometries. 
In the superhydrophobic case, the first peaks are more separated from the walls than in the  hydrophobic one. %, confirming a similar behavior. 
A similar behavior has been observed in Ref.~\cite{mar:ent17,Li:prb07}, where the authors studied a confinement between parallel walls 
at a similar separation (1.7\,nm). 
% the authors work with a system confined between two walls separated by a distance of 1.7 nm where bulk density is never recovered in systems with high confinement. 
In the hydrophobic case, due to the stronger water-wall interaction, 
 the first peaks appear closer to the walls, reaching a maximum density 
   of $\rho = 2.25$\,g/cm$^3$,  almost twice the maximum value reached for $\varepsilon_{co}^{sh}$. 
  This result implies that water is more structured close to the wall in the hydrophobic system than in the superhydrophobic one.

A similar effect is observed for the W-2 superhydrophobic case (see Fig. \ref{dens-walls})
%\ref{dens-walls}-c)
%hydrophobic (\ref{dens-walls}-d)
%systems.
%, where the superhydrophobic interaction avoids water to 
 % be near the surface (\ref{dens-walls}-c) while in the hydrophobic case, the water-wall interaction is stronger and two 
 % high and sharp peaks are shown in the density profile (\ref{dens-walls}-d). 
 As expected, in the middle of the simulation box, where the effect of the interface disappears, the system behaves as in the bulk phase 
 and the density profile resembles  the bulk 
 density value independently on the $\varepsilon_{CO}$ value used. 
%{\bf CV : COULD YOU DESCRIBE THE DIFFERENT PANELS a b c d OF THE PARALLEL WALLS?}

\begin{figure}
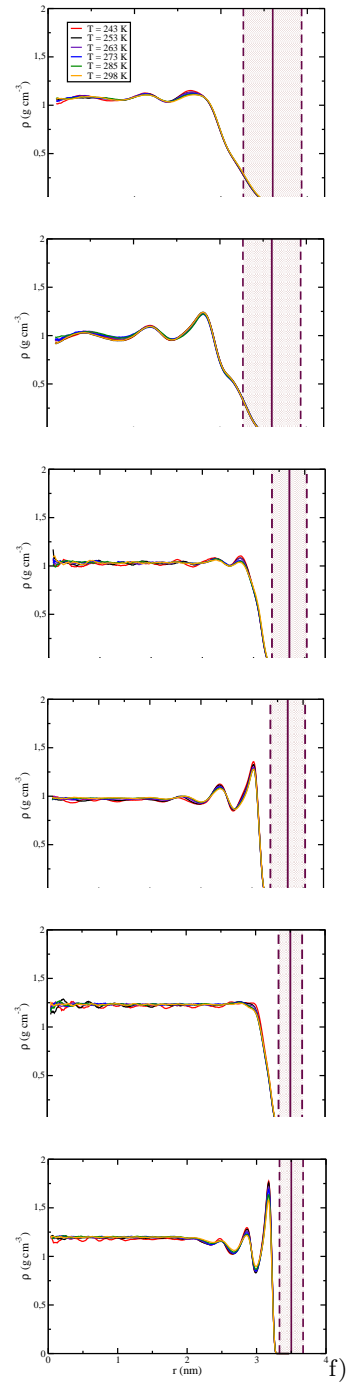

	\centering
	\includegraphics[height=2.95cm]{all_densprof_20x20_0047.eps}a)
	\includegraphics[height=2.95cm]{all-dens-prof-20x20-027-perfect.eps}b)
	\includegraphics[height=2.95cm]{all_densprof_35x35_0047.eps}c)
	\includegraphics[height=2.95cm]{all_densprof_35x35_027.eps}d)
	\includegraphics[height=2.94cm]{all_dens_prof_52x52_0047.eps}e)
	\includegraphics[height=2.94cm]{all_dens_52x52_027.eps}f)

	\caption{Density profiles  for water confined in CNT-1 (top panels), CNT-2 (middle panels)  and CNT-3 (bottom panels)  
		geometries at different temperatures (legend shown in figure a)). 
		Results for  the superhydrophobic system $\varepsilon_{co}^{sh}$ are on the left-hand side (a, c and e)
		whereas for the hydrophobic system are on the right-hand side 	  $\varepsilon_{co}^{h}$ ( b, d and f) .}
	\label{tubos}
\end{figure}

Figure~\ref{tubos} presents the   
density profiles obtained for water confined in CNT-1, CNT-2 and CNT-3  nanotubes  in a wide temperature range,  considering hydrophobic and superhydrophobic walls.
%are collected in Figure~\ref{tubos} representing the values for $\varepsilon_{co}^{sh}$ on the left and for $\varepsilon_{co}^{h}$ on the right the figure. 
%
When the interaction is superhydrophobic, 
changing the geometry of the confinement (from planar walls to cylinders) does not affect what was already observed for planar walls:
%, the effect of the  $\varepsilon_{co}$ is the same, 
%for superhydrophobic potential 
the peaks of the density profile are farther away from the  surface than in the hydrophobic case  
%more shifted  from the interface than for hydrophilic potential, 
while the peaks heights are higher in the latter case.
%and the peaks height is greater for the latter potential. 
Focusing on the CNT-1 system, when the interactions with the surface are hydrophobic (Fig.~\ref{tubos}-b), 
 the water density increases as approaching to the CNT surface reaching a density maximum of $\rho = 1.23$\,g/cm$^{3}$
   at $r = 0.90$\,nm at the highest temperature ($T = 298$\,K). 
When considering  superhydrophobic interactions (Fig.~\ref{tubos}-a), this peak is lower 
($\rho = 1.14$\,g/cm$^{3}$ at $r = 0.85$\,nm).  
%\textcolor{blue}{AL correction not implemented}
Interestingly, the highest peak of the  CNT-1's density profile, the one with  the largest curvature, presents 
 a shoulder that is closer to the carbon atoms than water confined in any other other system. 
This shoulder, originated by the curvature inducing a closer interaction between water  and  carbon, 
is the signature that the large curvature forces water to structure close to the hydrophobic nanotube.

Similarly, in  the CNT-2  a similar displacement of the maximum of the density profile can be detected:   the maximum in the density profile reaches a 
value of $\rho = 1.26$\,g/cm$^{3}$  at $r = 2.03$\,nm in the hydrophobic case (Fig.~\ref{tubos}-d) and of $\rho = 1.03$\,g/cm$^{3}$ 
at $r = 1.86$\,nm in the superhydrophobic case (Fig.~\ref{tubos}-c). 
Moreover, in the latter case,  the density profile slightly decreases with increasing temperature. 
%a shift of the density profile with respect to the temperature (decreasing by increasing the temperature), 
%negligible in the hydrophobic system as we observe in  Fig.~\ref{tubos}-d.
Also in  the CNT-3  a similar displacement of the maximum of the density profile can be detected:  
%Taking a look at the density profile corresponding to the CNT-3 system, one might conclude that
 the density  reaches a 
maximum at $\rho = 1.24$\,g/cm$^{3}$ and fluctuates around this value independently of the water-carbon interaction (see Fig.~\ref{tubos}-e and Fig.~\ref{tubos}-f).

As can be observed for all cases considered in Figs.~\ref{tubos} and \ref{walls}, the effect of the temperature on the density profiles is not very noticeable. 
% that there is just a slightly temperature dependence 
%on the density profile, independently the type of confinement of the degree of curvature.

As expected, the density profiles obtained for CNT-1 and CNT-2 are in excellent 
agreement with the results reported by K\"ohler {\it et al}.~\cite{koh:pccp17}  
for CNT (16,16) (at $\rho_1=0.93$\,g/cm$^{3}$  and $\rho_2=1.12$\,g/cm$^{3}$) 
and for CNT (30,30) (at $\rho_3=0.95$\,g/cm$^{3}$  and $\rho_4=1.15$\,g/cm$^{3}$) using TIP4P/2005 as water 
model.

\subsection{Hydrogen bonds profiles}

\begin{figure}
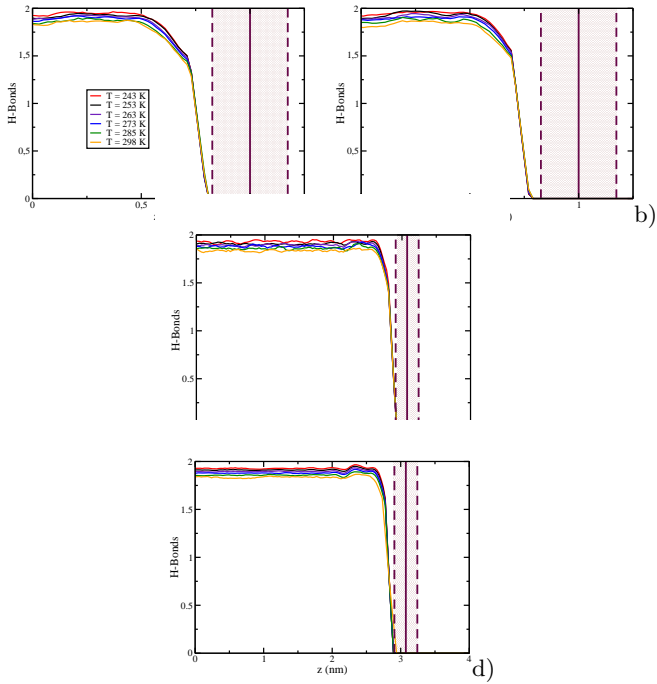

\centering
	\includegraphics[height=2.9cm]{hbonds-smallwalls-0047-true.eps}a)
        \includegraphics[height=2.9cm]{hbonds-smallwalls-027-true.eps}b)
	\includegraphics[height=2.9cm]{hbonds-5417-0047.eps}c)
	\includegraphics[height=2.9cm]{hbonds-bigwalls-027.eps}d)

\caption{Number of hydrogen bonds per molecule calculated for water confined 
between parallel walls: W-1 (top panels) and W-2 (bottom panels). Left-hand side (a and c) superhydrophobic  
 and right-hand side (b and d) hydrophobic surfaces.}
\label{hbondwall}
%\end{center}
\end{figure}

To further study the molecular structure of confined water, we analyzed how the average 
number of HBs of water molecules varies with respect to the distance from the 
confining surface.
%Other relevant property that reveals information on the molecular structure of 
%water inside the confinement is the hydrogen bonds (HB). 
%\textcolor{blue}{comentario de AL} It is well known that  a water molecule in bulk has 
%a specific orientation and number of first neighbours 
%A tetrahedron with a molecule of water inside de 
%centre and having 
%\pb{Two water molecules as acceptors and two as donors.  
%is the distribution for a molecule of water in bulk. 
%Therefore, the average number of HB for a molecule in bulk is around $2$. LAURENT: Those two sentences are not very clear, could you reformulate them? }   

A water molecule can be donor and acceptor of two hydrogen bonds. In such way, we chose one of the two possible roles of each molecule and we computed the HBs average using this criterion: therefore the average number of HB of a bulk molecule is around $2$.

\begin{figure}
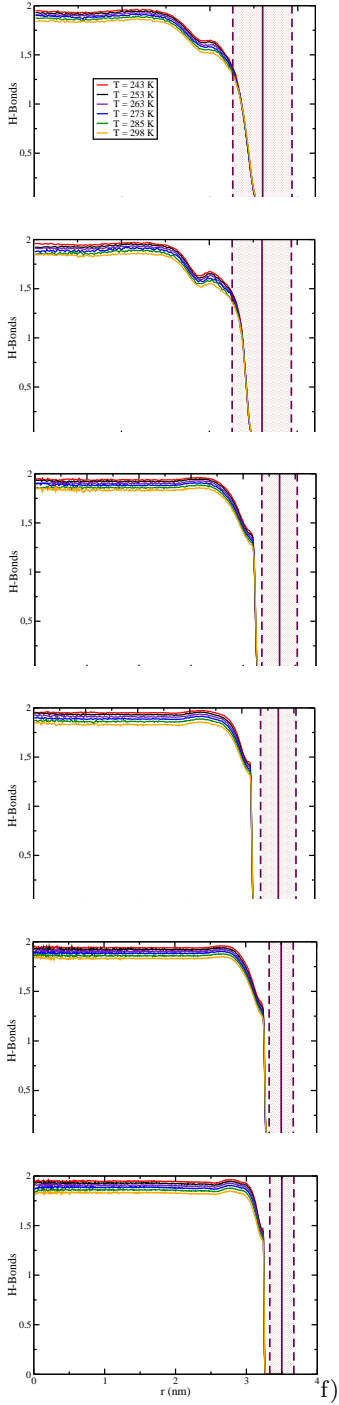

\centering
	\includegraphics[height=3.0cm]{all_sumhbonds_20x20_0047.eps}a)
	\includegraphics[height=3.0cm]{all_sumaahbond_20x20_027_final.eps}b)
	\includegraphics[height=3.0cm]{H-bonds_35x35_0047.eps}c)
	\includegraphics[height=3.0cm]{H-bonds_35x35_027.eps}d)
	\includegraphics[height=3.0cm]{all_sumas_hbond_52x52_0047.eps}e)
	\includegraphics[height=3.0cm]{all_hbonds_52x52_027.eps}f)

\caption{Number of hydrogen bonds per molecule calculated for water confined 
inside nanotubes: CNT-1 (top panels), CNT-2 (middle panels) and CNT-3 (bottom panels). Left-hand side (a, c and e) superhydrophobic  
 and right-hand side (b, d and f) hydrophobic surfaces.}
\label{hydrogen_bondsCNT}
\end{figure}

When water is confined, the number  of HBs per molecule decreases, especially close to the 
 (super)hydrophobic surface.  
%However, when a surface is present this distribution of impossible of keep it and the 
%
Figure~\ref{hbondwall} shows  the average number of hydrogen 
bonds   for confined water between two parallel walls (a and b for W-1- and c and d for W-2) within  a wide temperature range. 
Similarly,  Fig.~\ref{hydrogen_bondsCNT} represents the analogous property for a cylindrical 
confinement (a and b for CNT-1, c and d for CNT-2 and e and f for CNT-2). 
%Panels~\ref{hbondwall}-a and~\ref{hbondwall}-b show the hydrogen 
%bond profiles obtained at every temperature for W-1 applying 
%$\varepsilon_{co}^{sh}$ and $\varepsilon_{co}^{h}$ interaction respectively. 
%On the other hand,~\ref{hbondwall}-c and~\ref{hbondwall}-d plots represent the 
%HB profiles computed for W-2 system. 
As in bulk water, in every system  in Fig.~\ref{hbondwall} 
and  Fig.~\ref{hydrogen_bondsCNT}, we observe that the number of HBs decreases 
as the temperature increases, given that a higher temperature corresponds to a higher kinetic energy that 
leads to a higher bond breakage.
% there is a trend in which at higher the 
%temperature the number of HBs per molecule is decreased. 
%This could 
%be explained by  the increase of the temperature causes an increase of the 
%kinetic energy breaking the HBs. This behaviour is essentially 
%identical to bulk water.  

Close to the wall, regardless the type of confinement, 
the number of HBs decreases % \textcolor{blue}{comentario de AL}
until zero showing a region 
with a value less than 2, in good agreement with the Hydrogen bonds profile studied 
previously by Werder and coworkers.~\cite{wer:nanolett01} 
Due to the structuring of liquid water in the proximities of the surface (as shown in Figs. 3 and 4), 
more pronounced in the hydrophobic than in the superhydrophobic system, 
the  number of HBs presented is on average higher closer to the surface, 
especially for the hydrophobic systems. 

Focusing on CNT-1, we observe a shoulder (especially in the hydrophobic case) on the number of HBs near the surface
together with  a non-zero number of HBs inside the surface: 
these are both effects of large curvature of the system.
%On the other hand, panels \ref{hydrogen_bondsCNT}-c and 
%\ref{hydrogen_bondsCNT}-d show HB profiles for nanotube CNT-2 system. 
However, for both CNT-2 and CNT-3 systems the number of HBs 
is always around 2, being slightly higher before  approaching the surface. At that point the 
number of HBs dramatically drops to zero, due to the (super)hydrophobicity 
of the walls. 
%A slightly shoulder could be distinguish before a steep fall close to 
%the interface. 
%Finally, \ref{hydrogen_bondsCNT}-e and \ref{hydrogen_bondsCNT}-d plots, 
%represent the HB profile for CNT-3 system, the same shoulder near 
%the nanotube walls is observed as well as the pronounced decrease near the 
%carbon surface. 
%{\bf COULD YOU PLEASE DESCRIBE ALL PANELS a b c d OF FIGURE 7?}
In  the CNT-2 and CNT-3 geometries, curvature effects are almost negligible while the bulk 
contribution (towards the center of the nanotubes) increases.

\subsection{Diffusion Coefficient}

%\pb{LAURENT: Once Figs. 7, 8 and 9 are updated, we should check that everything below that point is consistent with the new figures. }

Having studied the structural properties of water under nano-confinements, we computed its transport properties 
for planar confinements (W-1 and W-2) and for two cylindrical confinements (CNT-1 and CNT-2).
In the latter case, we are not considering CNT-3 since this system behaves as the bulk. Therefore the CNT-2 system is enough to draw conclusions about  nanotubes with a large diameter.

We first compute the diffusion coefficient $D$ in a wide temperature range 
(243\,K - 298\,K) and for the different hydrophobicities of the surface.

As discussed in section~\ref{sec:sim_details}, the effect of the finite lateral size of the simulation box is simply dealt with by computing the mean square displacement of the particles after removing the motion of the center of mass; in confined systems, this provides a reasonable estimate of the intrinsic self-diffusion coefficient (i.e., in the limit of an infinite lateral box size). Therefore, the diffusion coefficients presented are the bare values measured from the mean square displacement.  

\begin{figure}
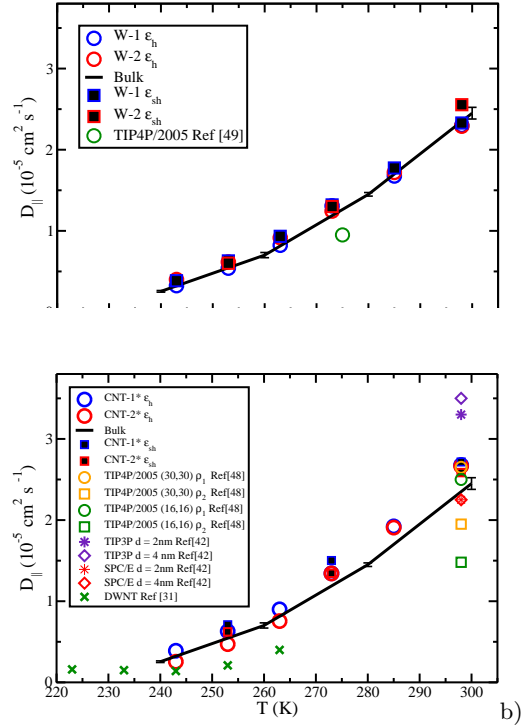

	\centering
	\includegraphics[width=0.75\linewidth]{MSD-walls-corregidos.eps}a)\\[2mm]
 \includegraphics[width=0.75\linewidth]{MSD-nanot-corregidos027.eps}b)
	\caption{Diffusion coefficient of water as a function of temperature when confined a) between hydrophobic and superhydrophobic walls, b) inside hydrophobic and superhydrophobic nanotubes. 
		The green empty circle in a) is for water confined between sheets (d = 1.4 nm) from Ref.~\cite{mar:ent17}. Bulk diffusion coefficient from Ref.~\cite{pab:jcp18} is represented by a black line. Stars and diamonds in panel b represent the diffusion inside nanotubes with a diameter of 2-4 nm  for TIP3P water (purple) and  SPC/E water (red) from Ref.~\cite{Ale:chemrev08}.
		Orange and green symbols represent results for water at densities $\rho$ = 0.95 g/cm$^3$ and $\rho$ = 1.15 g/cm$^3$ (circles and squares respectively)
		nanotubes with a chirality of (16,16) and (30,30) from~Ref.~\cite{koh:pccp17}.}
	\label{diff-correg}
\end{figure}

Results for water in planar confinement (W-1 and W-2 and the two different hydrophobicities) are presented in Fig.~\ref{diff-correg}-a, 
which shows  that the diffusion coefficient of  water confined under parallel 
walls is independent of the distance between walls (W-1 or W-2),
and is  similar to that of bulk water in  the same temperature range: 
 at low temperatures, the system presents slower diffusion.
 Moreover, the amount of hydrophobicity of the surface does not affect significantly the diffusion coefficient. 
In the same panel we also report the result from Ref.~\cite{mar:ent17}  
obtained for TIP4P/2005 water in a more stringent confinement.
The discrepancy with respect to our results might be due to the fact that water in Ref.~\cite{mar:ent17} 
is confined in a much tighter slab. This leads to an even more pronounced structuring of liquid water, 
that strongly affects its dynamic properties.

The results obtained for CNTs are presented in 
%The water diffusion obtained under cylindrical confinement is represented in 
Fig.~\ref{diff-correg}-b. 
Interestingly, the diffusion coefficient is affected by the curvature, 
being higher for the most curved surfaces (CNT-1 and CNT-2), and is barely affected by the 
hydrophobicity of the surface (being slightly higher in  the hydrophobic system).
As in the parallel walls confinements, the behavior of the diffusion coefficient with 
temperature is similar to the bulk. % as in the bulk, being the diffusion faster at higher temperatures and slower  at lower temperatures. 
%wall-confinement case, the diffusion coefficient for CNT systems with  temperature is as similar as bulk water.   
When $T \le 263$\,K, the values of the diffusion coefficient 
resemble those of bulk water independently on the curvature of the geometries and on their hydrophobicity. 
%\pb{No so clear to me: cf. in terms of relative difference? How would it look with a logarithmic scale for D? }
In contrast, at high temperature, the diffusion coefficient of confined water is  higher 
than the corresponding value for bulk water.

Focusing on $T = 298$\,K, where the effect is more dramatic, we compared our results with results from the literature.  
As already suggested in Ref.~\cite{liu:jcp14},  the diffusion coefficient of confined water is strongly affected by the water model,% chosen 
with TIP3P giving values too high with respect to SPC/E.\cite{Ale:chemrev08}
When comparing with TIP4P/2005 confined water at similar density and for similar chirality (orange symbols), 
we obtain results in perfect agreement with the literaturei.~\cite{koh:pccp17} 

Therefore, water  diffuses faster when confined in hydrophobic nanotubes with an effective diameter ranging from 2 nm up to about 4 nm (CNT-1 and CNT-2, respectively).

In order to unravel the microscopic origin of this behavior, we evaluate how the diffusion 
coefficient varies with the distance from the hydrophobic surface.  
\begin{figure}
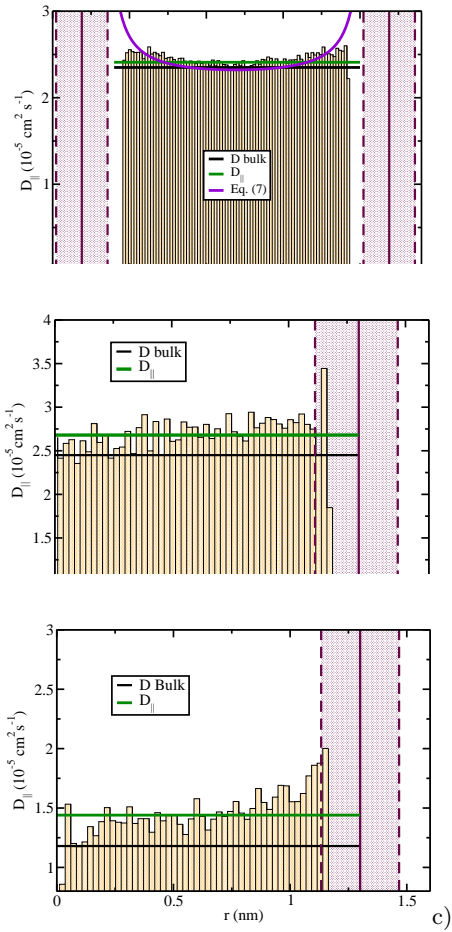

	\centering
	\includegraphics[height=4cm]{Dxy_vs_z_W1_027E_298K_v2.eps}a)
	\includegraphics[height=4.cm]{Dz_vs_r_CNT1_027E_298K.eps}b)
	\includegraphics[height=4.cm]{Dz_vs_r_CNT1_027E_273K.eps}c)
	\caption{Diffusion coefficient %of all molecules (normalised by the number of molecules and frames) represented 
		along the distance between walls or the radius of the nanotube for W-1 at  
		$T = 298$\,K (panel a), CNT-1 at  $T = 298$\,K (panel b) and at $T = 273$\,K  (panel c). 
		The green line represents the diffusion value reported in Fig.~\ref{diff-correg} at the same conditions and the black line 
		is the diffusion of bulk water from Ref.~\cite{pab:jcp18}.The purple line represents the diffusivity profile calculated via Eq. (\ref{eq:D_z_theo}). }
	\label{fig_D_vs_r}
\end{figure}

Figure~\ref{fig_D_vs_r} represents the diffusion coefficient of water molecules when confined: 
a) between parallel walls (W-1) at 298\,K and inside an hydrophobic nanotube (CNT-1) at 298\,K (b) and 
at 273\,K (c).  
Densities of water confined in those systems are similar, but
curvatures and structures are definitively different.  
Data for $D$ are plotted along the $XY$ direction for W-1  
(Fig.~\ref{fig_D_vs_r}-a) and the $Z$ direction for CNT-1 (Fig.~\ref{fig_D_vs_r}-b
and ~\ref{fig_D_vs_r}-c). 

For the slit pore at 298\,K (W-1, Fig.~\ref{fig_D_vs_r}-a), we observe that the average diffusion, computed with Eq. (2) (green line)
is practically the same as that for bulk water (continuous black line).
Figure~\ref{fig_D_vs_r}-a also shows the prediction of continuum hydrodynamics for the local diffusivity, Eq.~\eqref{eq:D_z_theo}. The shape of the theoretical curve matches well the measured values, except very close to the walls, where the diffusivity increment is lower than the continuum prediction. 
On the one hand, this could simply be because Eq.~\eqref{eq:D_z_theo} was derived for a distance to the walls $z\gg \sigma_\text{h}$, so that it would be outside its range of validity. 
On the other hand, the large quantitative discrepancy could be due to beyond-continuum effects, related in particular to the large changes in the (H-bond) structure highlighted in sections 3.1 and 3.2.

Note that in Eq.~\eqref{eq:D_z_theo}, the confined Stokes-Einstein viscosity, $\eta_\text{SE}$ was used, so that the average of Eq.~\eqref{eq:D_z_theo} corresponds to $\langle D_\| \rangle$. As a consequence, because $D_\| (z)$ is overestimated by Eq.~\eqref{eq:D_z_theo} close to the walls, it is also slightly underestimated in the middle of the slab, by  ca. 3\,\%  . This value provides an estimate of the typical error in the viscosity computed by the confined Stokes-Einstein method, where Eq.~\eqref{eq:D_z_theo} is assumed to be valid everywhere in the slab.

As shown in Fig.~\ref{fig_D_vs_r}-b and Fig.~\ref{fig_D_vs_r}-c, the 
curvature plays an important role in determining the diffusion of water, as previously suggested. 
~\cite{ala:chemphys11,fal:nanolett10}
When considering the effect of the curvature (CNT-1)
we observe that the average diffusion 
is higher  than the value for bulk  water at 298\,K  (Fig.~\ref{fig_D_vs_r}-a), and 273\,K  (Fig.~\ref{fig_D_vs_r}-c).
%Note that, the difference between 

\subsection{Viscosity}

\begin{figure}
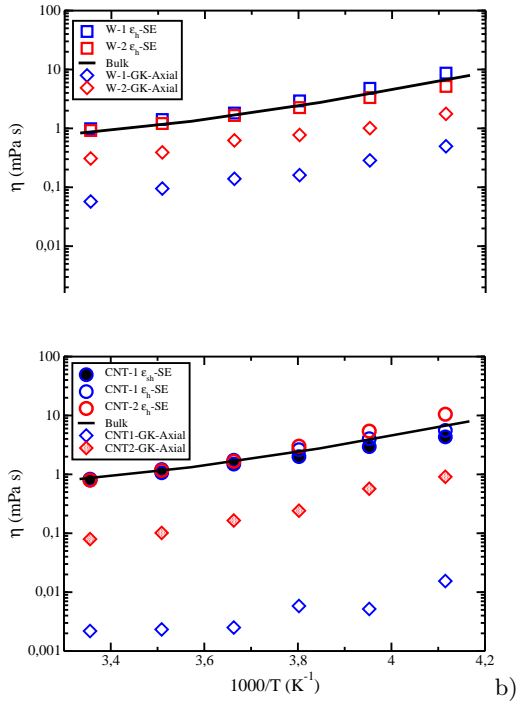

	\centering
	\includegraphics[width=0.75\linewidth]{COMP-GK-HIDRO-WALLS.eps}a)
	\includegraphics[width=0.75\linewidth]{COMP-GK-HIDRO-nanotub.eps}b)
%	a) \includegraphics[width=0.7\linewidth]{etha-T-walls}\\[2mm]
%	b) \includegraphics[width=0.7\linewidth]{etha-T-nanotubes}
	\caption{Viscosity $\eta$ as a function of temperature for water in slits (a) and in nanotubes (b). Blue color represents the smallest system in each case (W-1 or CNT-1) and red color represents W-2 or CNT-2. Black line shows the results reported in Ref.\cite{pab:jcp18} for bulk. The Green-Kubo estimate (diamonds) and the confined Stokes-Einstein estimate (squares and circles) are compared.}
	\label{fig:ViscoVsT}
\end{figure}

We now focus on the calculation of the viscosity, which we evaluate via two routes: the Green-Kubo expression, given in %(which gives information of both axial and radial viscosities), 
Eq.~\eqref{eq:visco} and the confined Stokes-Einstein expression, given by Eq.~\eqref{eq:eta_SE}, derived using continuum hydrodynamics.  %theoretical model for a confined system.
Concerning the Green-Kubo approach, as  explained in Sec.~\ref{sec:sim_details}, we use only some components of the pressure tensor, depending on the geometry of the confinement. Therefore, care must be taken when comparing to literature results obtained for bulk water, where all pressure tensor's components are taken into account. 
% Concerning the confined Stokes-Einstein approach, as  explained in Sec.~\ref{sec:sim_details}, having numerically computed the diffusion coefficient, we derive the viscosity Eq.~\eqref{eq:eta_SE}.
%Details on the calculations of the viscosity for W and CNT systems are reported in 
%Fig.~\ref{walls-visco4} of Appendix 3.

%%%%%%%%%%%%%%%%%%%%%%%%%

%\noindent \large{\textbf{Appendix 3: Viscosity}}
%\label{app:visco}

We focused on the hydrophobic case,  computing the viscosity for both slit pore and nanotube systems in a wide temperature range (243-298\,K).
Our results are presented in 
%  while for superhydrophobic potential it 
%was calculated for the most relevant cases to compare the effect of hydrophobicity. 
%In order to clarify the comparison of the results, we have collected the most 
%relevant viscosity data on
%Table~\ref{tab:visco} and  
Fig.~\ref{fig:ViscoVsT}-a for the hydrophobic slits and Fig.~\ref{fig:ViscoVsT}-b for CNT-1 and CNT-2 nanotubes. 
Consistently with the diffusivity, the confined Stokes-Einstein viscosity reported in panel a for a slit pore shows a very good agreement with the bulk values (continuous black line) independently on the system size (W-1 blue and W-2 red squares). %This feature has been already observed when measuring the diffusion coefficient. 
In contrast, the Green-Kubo axial viscosity (empty diamonds) depends strongly on system size (with the largest W-2 system closer to the bulk values than the W-1 one). Note that similar results for the viscosity have been obtained using the GK expression by ref. \cite{koh:pccp17} for a similar size CNT (same density and hydrophobicity).

Similar results are obtained for nanotubes (panel b): $\eta_\text{SE}$ is always very close to the bulk values (continuous black line) independently of the system size (CNT-1 blue and CNT-2 red circles) and on hydrophobicity ($\varepsilon_{co}^{h}$ empty and $\varepsilon_{co}^{sh} $ filled symbols). 
%As expected, we detect 
A small discrepancy appears at low temperatures, which could be related to the failure of the Stokes-Einstein relation in supercooled water. 
In contrast, a clear system size dependence is detected for the Green-Kubo axial viscosity (empty diamonds); for the largest CNT-2 system viscosity values are closer to those of the bulk when compared with those of the CNT-1 one.
%Here also, the viscosity increases as the temperature decreases in all cases.

%%%%%%%%%%%

To conclude, while viscosity computed with the Green-Kubo formula (applied for anisotropic and confined system) strongly differs from the bulk, viscosity computed with the confined Stokes-Einstein relation is not so affected by confinement, independently on its geometry.

\section{Summary and conclusions}

In this work we present a thorough study of TIP4P/2005 water confined in different hydrophobic nanostructures  
%We also used two carbon-water interactions : hydrophobic 
%($\varepsilon_{co}^{h} = 0.2703$\,kJ/mol) and superhydrophobic 
%($\varepsilon_{co}^{sh} = 0.0470$\,kJ/mol) and we evaluated every system at a 
within a wide  temperature range.
Properties such as density profiles, HBs per molecule, diffusion and 
viscosity were obtained using MD simulation.

Studying the density profiles, we concluded that water density approaches the  bulk value in the middle of  
either the slit pore or the nanotube, while more structured water 
  (highest peaks)  is observed near the walls. 
  We detect differences of the density profiles depending on the amount of hydrophobicity considered: 
  water seems more structured in the hydrophobic cases rather than in the superhydrophobic ones. 
  %  It was also 
%demonstrated a significative difference between the profiles calculated for hydrophobic, where water gets more 
%structured, presenting high and thin peaks;  and superhydrophobic interactions, where molecules are placed 
%further the walls and the density peaks are lower than the ones computed in the hydrophobic case.}
%\textcolor{red}{When performing a  HB analysis, we find the latter strongly correlated to the density profile.}
%A HBs analysis was also performed, revealing a concordance with the density profile, 
As expected, the number of HBs in bulk water is reached
in the middle of either the slit pore or the nanotube;  while near the surface a sharp decrease to 
zero is observed, given that those molecules have a smaller number of water neighbours, thus forming less bonds. 
Interestingly, in the CNT-1 system, we observe a shoulder that we attribute to 
curvature effects. 
% however deeper studies in this lines are required.   

To study diffusivity, we treated effects of the finite lateral size of the simulation box by 
simply computing the mean squared displacement of the molecules after removing the motion of the liquid center of mass. 
Interestingly, when confined in nanometer size nanotubes, water tends to diffuse faster than in bulk (CNT-1 and CNT-2). Considering that CNT-1 has the highest curvature, we  conclude that curvature is an important parameter that can be handled to  produce an increment of the diffusion while confining water, especially at lower temperatures.
We showed that the confinement of water induces a significant  
difference between the diffusion of the molecules in the center of the 
systems and the molecules close to the interface as shown in figure \ref{fig_D_vs_r}, in qualitative agreement with the prediction of continuum hydrodynamics. The quantitative discrepancy could be either due to an approximation in the derivation of the analytical prediction, or to non-continuum effects, consistently with the large structural changes close to the walls.  
%This effect occurs for both hydrophobic and superhydrophobic types of walls.

We then compared the two methods to evaluate viscosity. 
First, we used a Green-Kubo expression (whose applicability in confined heterogeneous systems is not guaranteed), considering only some components of the pressure tensor for computing axial viscosity as commented in Sec.~2.3. 
%. 
Then, we derived a confined Stokes-Einstein relation, taking into account the influence of the (stress-free) confining walls, based on a previous expression derived for no-slip walls \cite{Simonnin2017}. Viscosity can also be estimated through this confined Stokes-Einstein relation, although this expression could fail at low temperatures, similarly to its bulk counterpart. 
We found that the two estimates differ dramatically, with the confined Stokes-Einstein estimate globally remaining closer to the bulk viscosity, and being more consistent with what we observed for diffusion (by construction). 
Since both methods have their shortcomings, here we only would like to suggest that measuring viscosity in a confined system is a delicate business, and that different methods may provide dramatically different results, which may not be easily related to the standard, experimental definition of viscosity. To conclude, care must be taken when computing viscosity in inhomogeneous and anisotropic confined systems. In terms of diffusion and viscosity, we do not see any dynamical signature of any liquid-liquid transition. At least within the chosen temperature range and nanotube diameter/plane geometry. Further work is needed to unravel the features of water under ultra-confinement, where water could undergo structural and/or dynamical transitions.~\cite{fum:sci18,cai:jcp19,che:jpc18,dix:jcc18}

\section*{Acknowledgements}
A.Z. thanks CONACYT for the PhD scholarship. M.A.G. acknowledges support by Ayuda Juan de la Cierva-Incorporación (IJCI-2016-27497) from Ministerio de Ciencia, innovaci\'on y universidades (Spain).This work was partially supported by CNRS (France) through a PICS program. L.J. acknowledges support from Institut Universitaire de France. C.V. acknowledges funding from Grant FIS2016-78847-P of the MINECO and together with I.L.M. and M.A.C.M. the UCM/Santander PR26/16-10B-2. The authors acknowledge the computer resources and technical assistance provided by Marenostrum (RES).

\appendix

\section{Contact angle measurements}

\begin{figure}
	\centering	
%	a) \includegraphics[height=1.9cm]{figure-hidrophil-wht1.eps}
%	\hspace{4mm} b) \includegraphics[height=1.9cm]{figure-hidrophob-wht.eps}
	      a) \includegraphics[height=1.9cm]{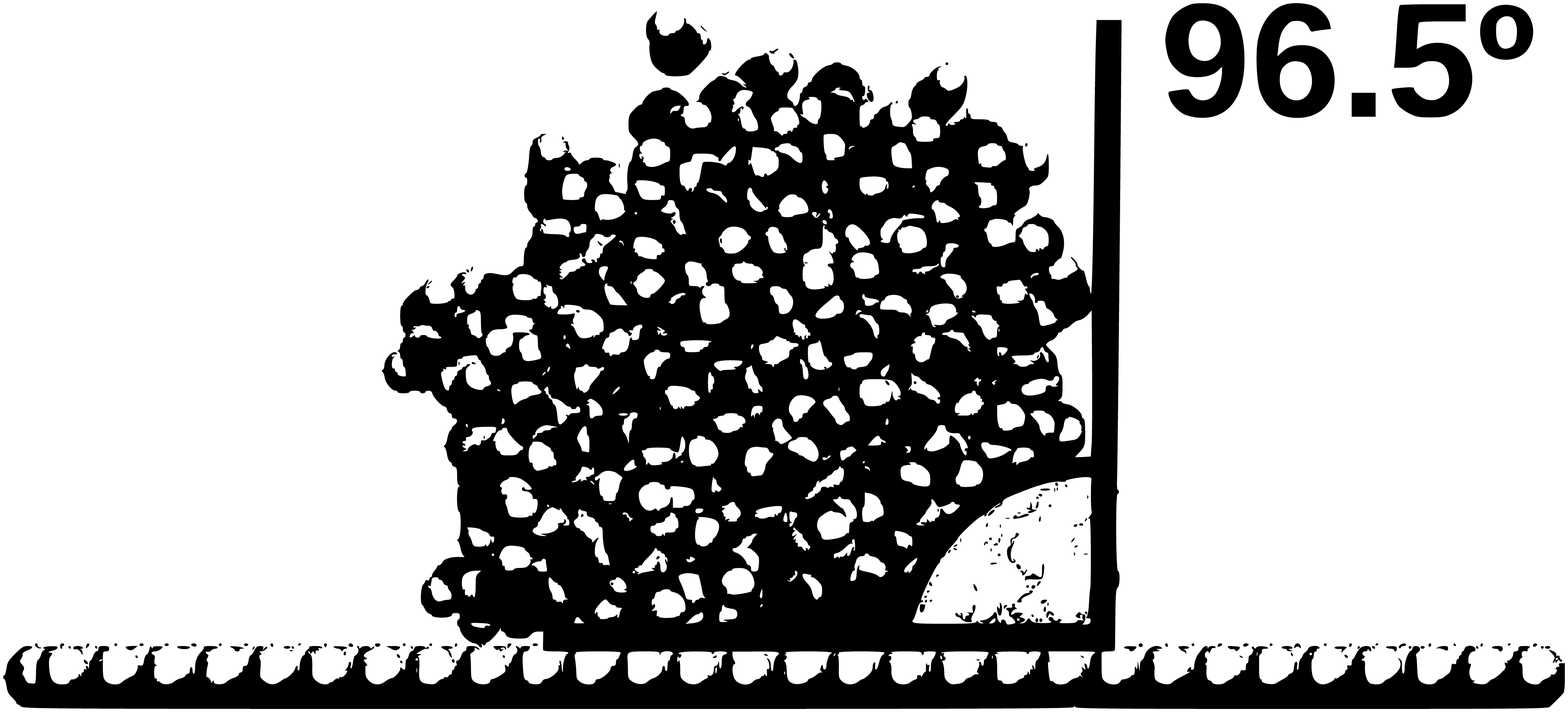}
              b) \includegraphics[height=1.9cm]{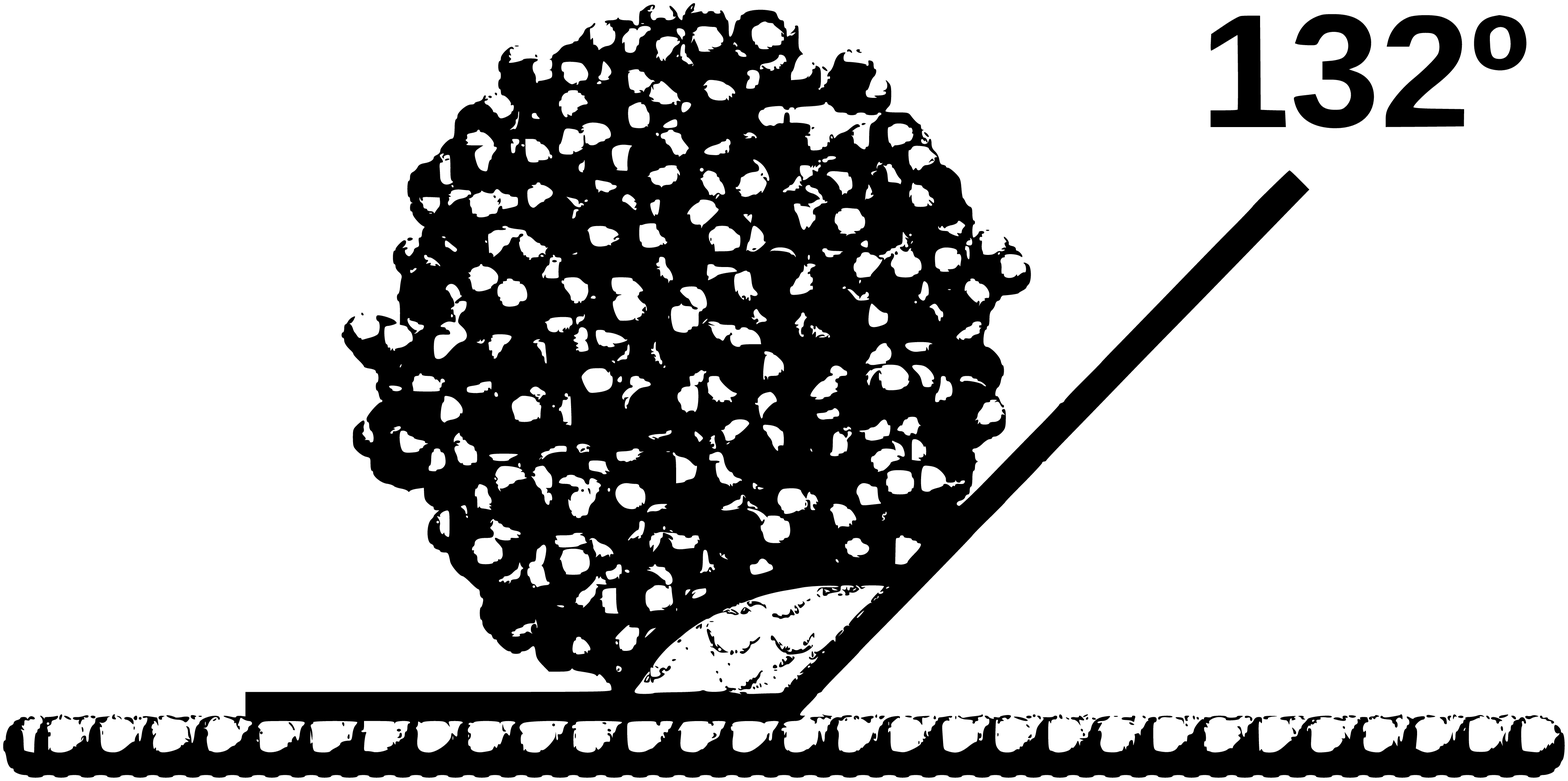}
	\caption{Snapshots representing liquid droplets on a flat surface. Carbon atoms are represented by cyan spheres and the water molecules are red and white (red for oxygen  and  white for   hydrogen). 
		(a) Hydrophobic water-wall interaction using $\varepsilon_{co}^{h}$ (b) Superhydrophobic water-wall interaction using $\varepsilon_{co}^{sh}$. 
		The numbers indicate the contact angles resulting from our calculations. }
	\label{drop1}
\end{figure}

In order to quantify the amount of hydrophobicity, we compute the contact angle of a liquid droplet located on a planar surface.
As in an experimental drop-shape analysis ~\cite{gui:jce10}, we prepare a  liquid droplet containing 200 water molecules and  locate it on top of a rigid and flat carbon surface.
Depending on whether the surface is hydrophobic (h) or superhydrophobic (sh), the surface will be "partially wet"  as we can see in Fig.~\ref{drop1}-a or dry  as in Fig.~\ref{drop1}-b.

%%{\bf CV FIGURE 1) NO PERSPECTIVE 2) CONTACT ANGLE AS IN YOUNG'S EQUATION}

Next, we estimated the contact angle between the droplet and the surface. 
After having equilibrated the system for 2~ns, 
we measured the contact angle by projecting on the $X-Y$ plane the water molecules and of the carbon atoms coordinates: 
%of the water droplet on the carbon surface by mapping out the coordinates of each moLecules on the $X-Y$ plane. 
defining the line formed by the outermost molecules in the droplet with the molecule in contact with the layer,
and a line defined on the plane of the carbon atoms, then we %and analyse them  using a graphic software \textcolor{green}{(xfig)} \textcolor{blue}{WHICH SOFTWARE? WE NEED TO CITE IT} 
computed the contact angle among them.
To get more statistics, we have used the coordinates of all molecules for 10 independent trajectories at same temperature, averaging the final result.

With this method, we estimated the contact angle for the hydrophobic case to be 
around 96.5$^{\circ}$ which indicates a slight hydrophobicity of the surface, similarly to the one computed by 
Hummer et al.\cite{hum:nat01}, 
Moskowitz et al.,\cite{mos:jcp14} and Kohler et al.\cite{koh:pccp17} considering the same interaction strength  ($\varepsilon$ = 0.2703 kJ/mol). 
In the superhydrophobic system, 
where $\varepsilon_{co}^{sh}$  is 0.047~kJ/mol, we obtained a contact angle of $\approx$ 132$^{\circ}$, much larger than 90$^{\circ}$.

\section{Confined Stokes-Einstein relation}
\label{sec:confinedSE}

Here we consider a liquid confined in an infinite slit pore of height $d$, assuming a perfect slip (stress-free) boundary condition at the walls, and a homogeneous and isotropic viscosity $\eta$. We derive an expression for the average self-diffusion coefficient parallel to the walls. To that aim, we follow the approach described in Ref.~\citenum{Simonnin2017}, with the only difference that we consider stress-free walls. 

We define the vertical position $z$ by the position of the hydrodynamic shear planes at $z = 0$ and $d$. The local diffusivity $D_{\|}(z)$ can be related to the local mobility $\mu_{\|}(z)$ through Einstein relation \cite{Einstein1905}: $D_{\|}(z) = k_\text{B}T \mu_{\|}(z)$. The mobility can then be computed within the framework of continuum hydrodynamics. 
In bulk, the standard Stokes prediction $1/\mu_{\|}^\text{bulk} = 3 \pi \eta \sigma_\text{h}$ %, with $\eta$ the viscosity, 
can be used to define the effective hydrodynamic diameter of the particles $\sigma_\text{h}$, see Table~\ref{hydrod}.  
In confinement, the mobility is modified due to friction on the two walls. When the distance to the walls is large as compared to the particle radius, one can write \cite{Saugey2005,Lauga2005,Joly2006a,Simonnin2017}:  
\begin{equation} 
\frac{1}{\mu_{\|}(z)} = 3 \pi \eta \sigma_\text{h} \left( \frac{1}{1+\frac{3}{8} \frac{\sigma_\text{h}}{2z+\sigma_\text{h}}}+\frac{1}{1+\frac{3}{8} \frac{\sigma_\text{h}}{2(d-z)+\sigma_\text{h}}}-1 \right) . 
\end{equation}
The local diffusivity is accordingly: 
\begin{equation} \label{eq:D_z_theo}
D_{\|}(z) = \frac{k_\text{B}T}{3 \pi \eta \sigma_\text{h}} \left( \frac{1}{1+\frac{3}{8} \frac{\sigma_\text{h}}{2z+\sigma_\text{h}}}+\frac{1}{1+\frac{3}{8} \frac{\sigma_\text{h}}{2(d-z)+\sigma_\text{h}}}-1 \right)^{-1} . 
\end{equation}
The average diffusivity is then given by: 
\begin{equation}
	\langle D_{\|} \rangle = \frac{1}{d} \int_{0}^{d} D_{\|}(z)\ \mathrm{d}z . 
\end{equation}  
The full solution is rather cumbersome, but in large pores ($d \gg \sigma_\text{h}$), it can be approximated by: 
\begin{equation}
\langle D_{\|} \rangle \approx \frac{k_\text{B}T}{3 \pi \eta \sigma_\text{h}} \left[1 + \frac{3}{8} \frac{\sigma_\text{h}}{d} \ln  \left( \frac{2d}{\sigma_\text{h}} \right) \right] . 
% + \mathcal{O} \left\{ \frac{\sigma_\text{h}^2}{d^2} \right\}
\end{equation}
In the main text, we will refer to this expression as the confined Stokes-Einstein relation, and use it to compute the liquid viscosity from its diffusion coefficient. In practice, we will use the effective wall position as the origin, and the effective wall distance for $d$ (see section 3.1 of the main text).

\begin{table}
	\begin{center}
	\caption{Bulk diffusivity $D_\text{bulk}$, bulk viscosity $\eta_\text{bulk}$ and effective hydrodynamic diameter $\sigma_\text{h} = k_\text{B}T/(3\pi \eta_\text{bulk} D_\text{bulk})$ as a function of temperature for a density of $999.26$\,kg/m$^3$. Data extrapolated from Ref.~\citenum{pab:jcp18} (we fitted the numerical data with Vogel-Tammann-Fulcher laws).}
	\label{hydrod}
	\begin{tabular}{|c|c|c|c|}
   \hline 
   \hline 
	   $T$ (K) & $D_\text{bulk}$ ($\times 10^{-9}$\,m$^2$/s) & $\eta_\text{bulk}$ (mPa\,s) & $\sigma_\text{h}$ (nm) \\
		\hline 
		298 & 2.35 & 0.849 & 0.219 \\
		285 & 1.67 & 1.18 & 0.212 \\
		273 & 1.14 & 1.71 & 0.205\\
		263 & 0.791 & 2.47 & 0.197 \\
		253 & 0.513 & 3.81 & 0.190 \\
		243 & 0.306 & 6.44 & 0.181 \\
   \hline 
		   \hline 
\end{tabular}
	\end{center}
\end{table}

%\noindent \large{\textbf{Appendix 4: Experimental NMR measurements }}
\section{Experimental NMR measurements}

In order to confirm our numerical results, we have attempted to  measure the diffusion coefficient inside the nanotubes by means of Nuclear Magnetic Resonance,  NMR, 
following the protocol described in Ref. \cite{liu:langmuir14},  that uses NMR DOSY (Diffussion Ordered SpectroscopY) experiments to experimentally determined the diffusion coefficients.

To start with, we prepare a suspension of single walled CNTs in water. 
We located a suspension of %\textcolor{blue}{comentario AL} 
300 $\mu$ l distilled water and 
1 mg  single walled CNTs with open tips at both ends (provided by Ionic liquid Technologies, O-SWCNT diameter=$1-2$nm, length=$1-3\mu$ m) in a $5$ mm  NMR tube. 
To properly disperse the nanotubes in water, we sonicate the suspension for 6 hours. 

Next, we have inserted in the sample  a coaxial inner tube containing CD3CN needed as an external lock solvent for the NMR experiment. 
We acquired $^1$H 1D NMR experiments on a Bruker Advance 500 MHz spectrometer with 16 scans at the following temperatures:  293 K, 278 K, 263 K and 253 K. 
Next, we performed NMR DOSY (Diffusion Ordered SpectroscopY) experiments  at 500 MHz with the standard Bruker DOSY protocol (stebpgp1s), collecting 
thirty two 1D $^1$H  spectra   with a gradient duration of $\delta$ = 1 ms and an echo delay of $\Delta$=100 ms.

\begin{figure} 
\begin{center}
	\includegraphics[height=3.5cm]{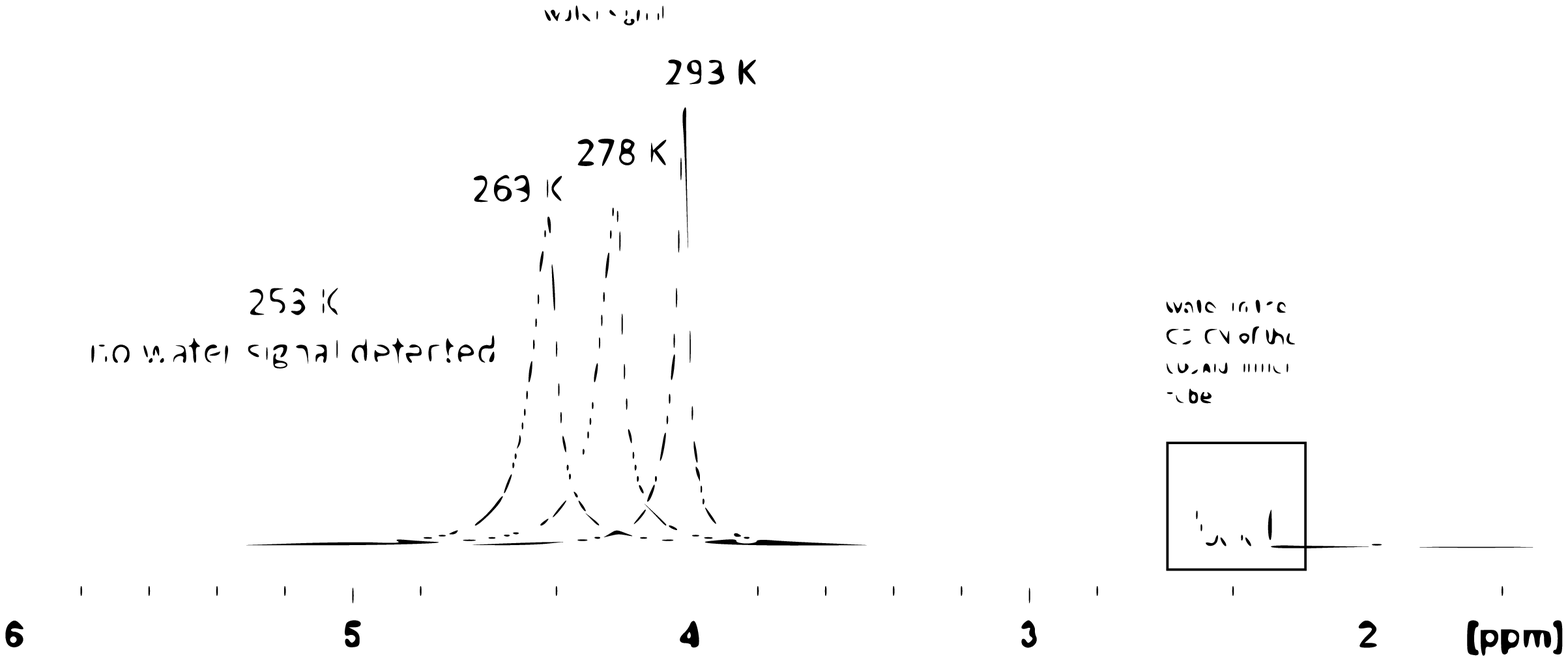}
	\caption{$^{1}$H spectra of the CNTs-water suspension  at 293 K (black curve), 278~K (red curve), 263~K (dark blue) and 253~K (green). \label{NMR1}}
\end{center}
\end{figure}

NMR spectra are shown in Fig. \ref{NMR1}. 
As expected,  the water signal shifts to the left for lower temperatures.

The $^1$H  NMR signal at 263 K displays line broadening (as its  signal in the DOSY experiment), signature of the fact that 
 the sample contains a mixture of bulk and frozen water ($log(K_D)= -(9.02-9.14)$ thus $K_D = (9.5*10^{-10}-7.2*10^{-10})$ $m^2/s$).
 No  signal has been  detected at the lowest temperature of 253 K due to the 
crystallization of the entire suspension.  
Interestingly, at every temperature we always noticed on the right a low peak that we attribute to a small amount of water 
present in the CD3CN capillary.

\begin{figure}
\begin{center}
	\includegraphics[height=3.5cm]{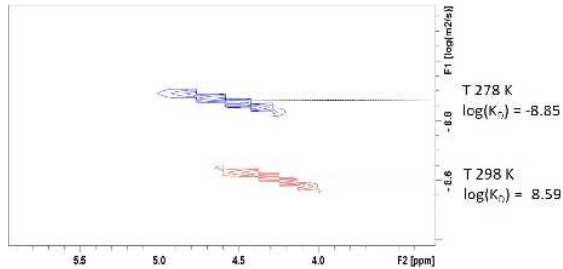}
	\caption{DOSY NMR experiments acquired at 298 K (red colour) and 278 K (blue colour) overlaid, where 
	$log(K_D)= -8.85$ (at 278K) and $log(K_D)= -8.59$ (at 298K).	
	\label{DOSY}}
\end{center}
\end{figure}

Results obtained from the NMR DOSY experiments are shown in  Fig.\ref{DOSY}.   

As reported in Fig \ref{NMR1}, the suspension is liquid at temperatures down to 263K. 
As shown in Table \ref{dosytable}, 
the obtained values of the diffusion coefficient 
at 298 K, 293 K and 278 K ($log(K_D)$= -8.59, -8.64, -8.85, respectively) are consistent with the reported values for bulk water 
presented in the literature. %\textcolor{blue}{comentario de AL} 

\begin{table}[h!]
\begin{center}
\begin{tabular}{|c|c|c|}
\hline
\hline
T (K) & D(cm$^{2}$ s$^{-1}$)  & D(cm$^{2}$ s$^{-1}$) Literature \\
\hline
\hline
298 & 2.57 $\times$ 10$^{-5}$&   2.29 $\times$ 10$^{-5}$ Ref.\cite{holz:pccp00} \\
\hline
 293 & 2.29 $\times$ 10$^{-5}$&   2.02 $\times$ 10$^{-5}$ Ref.\cite{holz:pccp00}  \\
\hline
 278 & 1.41 $\times$ 10$^{-5}$ &  1.31 $\times$ 10$^{-5}$  Ref.\cite{mil:jpc73}  \\
\hline
\hline
\end{tabular}
\caption{Comparison between experimental and reported bulk water diffusion coefficients
	\label{dosytable}}
\end{center}
\end{table}

%%%%%%%%%%%%%%%%%%%%%%%%%%%%%

%\textcolor{red}{We show the spectra, and detect that the peaks has to move towards the right when lowering the temperature (isn't it)}

%{\bf CV METER DETALLES EXPERIMENTALES AQUI}
%\begin{itemize}
%\item {\textcolor{red}{Describe NMR apparatus}}
%\item {\textcolor{red}{describe nanotubes features (info from Marta's email)}}
%\item {\textcolor{red}{describe sample preparation, chosen concentration, sonication,....}}
%\end{itemize}

Our experimental results for bulk water are represented together with our numerical results for water confined in the hydrophobic 
$CNT1$ and $CNT2$ nanotube and with the results presented in reference \cite{liu:langmuir14}.

%\section{\textcolor{red}{Experimental Measurements}}

\begin{figure}
\begin{center}
 \includegraphics[height=4.0cm]{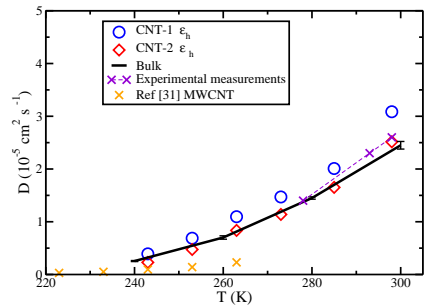}
\caption{Diffusion coefficient of water confined within $CNT1$ and $CNT2$ nanotubes (red and blue symbols). The purple symbols are results from our experiments while the orange ones from 
Ref. \cite{liu:langmuir14}. The solid black line represents the expected values for bulk water.}
\label{D_exp}
\end{center}
\end{figure}

Differently from us, Liu et al. \cite{liu:langmuir14} detected the $^1$H   water NMR signal at temperatures below 263 K:  
 the authors attributed this signal to endohedral mobile water inside the nanotubes. 
 One possible difference between the two experiments is 
the amount of confined water, that could not be enough  for NMR detection in our experimental conditions. 
Even though the diameter of the nanotubes is similar in both experiments ($1-2$ nm vs  $2.3$ nm  in \cite{liu:langmuir14}) 
their length is smaller in our experiments ($1-3$ $\mu$m vs $12-18$ $\mu$m in \cite{liu:langmuir14}). 
Another difference is that we used single wall, while the authors of  \cite{liu:langmuir14} used multi walled nanotubes. 
%However, we are not sure how this could affect the NMR measurements. 

\bibliographystyle{ieeetr}
%\bibliography{./bibliographyAZ,libraryLAURENT}

%\section{Bibliography}

%%\bibliography{library}

%%%%%%%%%%%%%%%%%%%%%%%%%%%%%%%%%%%%%%%%%%%%%%%%%%%%%%%%%%%%%%%%%%%%%%%%%%%%%%%%%%%%
%----------------------------------------------------------------------------------%
%%%%%%%%%%%%%%%%%%%%%%%%%%%%%%%%%%%%%%%%%%%%%%%%%%%%%%%%%%%%%%%%%%%%%%%%%%%%%%%%%%%%

\end{document}